\documentclass[pra,twocolumn,amsmath,superscriptaddress]{revtex4}

\usepackage{amsfonts}
\usepackage{mathrsfs}
\usepackage{amssymb}
\usepackage{amsthm}
\usepackage{pifont}
\usepackage{epsfig}
\usepackage{graphicx}
\usepackage{subfigure}
\usepackage{dsfont}
\usepackage{amsbsy}
\usepackage{mathrsfs}
\usepackage{amscd}
\usepackage{braket}
\usepackage{array}
\usepackage{bm}
\usepackage{xcolor}
\usepackage{makecell}
\usepackage{multirow}
\usepackage[hidelinks]{hyperref}

\newtheorem{theorem}{Theorem}

\newtheorem{corollary}{Corollary}
\newtheorem{lemma}{Lemma}
\newtheorem{example}{Example}

\begin{document}
\title{\large\bf Quantum separability criteria from bipartite systems to multipartite systems based on generalized Bloch representation}

\author{Linwei Li}
\affiliation{School of Mathematical and Sciences, Harbin Engineering University, Harbin 150001, China}

\author{Chunlin Yang}
\affiliation{School of Mathematical and Sciences, Harbin Engineering University, Harbin 150001, China}

\author{Hongmei Yao}
\email{hongmeiyao@163.com}
\affiliation{School of Mathematical and Sciences, Harbin Engineering University, Harbin 150001, China}

\author{Aimin Xu}
\affiliation{School of Mathematical and Sciences, Qufu Normal University, Qufu 273165, China}

\author{Zhaobing Fan}
\affiliation{School of Mathematical and Sciences, Harbin Engineering University, Harbin 150001, China}

\author{Shaoming Fei}
\email{feishm@cnu.edu.cn}
\affiliation{School of Mathematical Sciences, Capital Normal University, Beijing 100048, China}

%
%
%
%
%
%
%
%

\begin{abstract}
    Quantum entanglement serves as a fundamental resource in quantum information theory. This paper presents a comprehensive framework of separability criteria for detecting bipartite and multipartite entanglements. We construct a novel parameterized extended correlation tensor via the generalized Bloch representation under an arbitrary orthogonal basis, which improves the performance of entanglement detection. Moreover, we employ the generalized matrix unfolding to generalize the extended correlation tensor construction to multipartite systems, obtaining separability criteria for multipartite entanglement. Detailed examples demonstrate that our separability criteria exhibit enhanced capability in detecting entanglement.
\end{abstract}

\maketitle

\section{Introduction}
Quantum entanglement, a cornerstone of quantum information processing~\cite{nielsen2010quantum, horodecki2009quantum, huo2026sequential, li2025temporal}, underpins the theoretical foundations and practical applications of quantum computing~\cite{divincenzo1995quantum}, quantum communication~\cite{bennett1993teleporting, xu2020secure, xing2024teleportation, li2025communication}, quantum cryptography~\cite{ekert1991quantum}, and quantum simulation~\cite{lioyd1996universal}. Distinguishing entangled quantum states from separable ones is essential for ensuring the effectiveness and security of these quantum technologies. 

Separability criteria provide crucial tools in entanglement detection. Numerous separability criteria have been proposed, including the positive map criterion~\cite{peres1996separability, horodecki1996separability, li2024quantum}, computable cross-norm or realignment criteria~\cite{rudolph2003some, rudolph2005further, chen2002matrix, sun2025separability},  as well as criteria based on covariance matrices~\cite{guhne2007covariance} and Bloch representation~\cite{devicente2007separability}. While highly effective for bipartite systems, the extension of these criteria to multipartite mixed states remains challenging, primarily due to the intricate and richer entanglement structures in such systems. 
To address this complexity, tensor theory~\cite{bu2015minimum, yao2015tensors, yao2016singular, sun2016moore, sun2018generalized} has emerged as a powerful tool for characterizing correlations in multipartite systems. This approach captures the global properties of quantum states through correlation tensors, moving beyond local attributes and offering new perspectives for separability analysis.

Significant progress has been made in developing separability criteria based on correlation tensors. For instance, de Vicente~\cite{devicente2007separability} established an upper bound on the trace norm of the correlation tensor for separable bipartite states by using the Bloch representation. Shen et al.~\cite{shen2016improved} introduced parameters to an extended correlation tensor, deriving new bounds for both bipartite and multipartite states. Subsequent works have explored various bases for the Bloch representation, such as the Heisenberg–Weyl basis~\cite{chang2018separability,zhao2022detectionofmultipartite}, the canonical basis~\cite{sarbicki2020family}, and the Weyl basis~\cite{huang2022separability}, leading to a diversity of separability criteria. More recently, Zhu et al.~\cite{zhu2023family} constructed two novel matrices based on the Bloch representation to propose a separability criterion for bipartite systems. Huang et al.~\cite{huang2024unifying} further advanced this line of research by adopting an arbitrary orthogonal basis, presenting a unified form of extended correlation tensors and a corresponding separability criterion for bipartite states.
Despite these advances, existing Bloch‑representation‑based criteria vary in their tensor constructions and basis choices. 
A unified framework that is naturally applicable to bipartite and multipartite systems remains an interesting and challenging problem.

In this work, we propose such a framework based on a novel parameterized extended correlation tensor. 
By introducing independent vector parameters $\bm{u},\bm{v},\bm{\alpha},\bm{\beta}$ into the generalized Bloch representation under an arbitrary orthogonal basis, our bipartite criterion offers enhanced detection power and, with specific parameter choices, reduces to several existing ones. 
To extend this framework to multipartite systems, we employ the generalized matricization~\cite{brett2006algorithm, kolda2009tensor}, which flattens multiple dimensions of a tensor simultaneously into the rows or columns of the resulting matrix. 
While matricization has been used in multipartite separability studies, 
our key contribution is to combine it with the parameterized extended correlation tensor and provide criteria for detecting multipartite entanglement and genuine multipartite entanglement. 

The paper is structured as follows. Sect.~\ref{section: preliminaries} introduces some notations and definitions. Sect.~\ref{section: bipartite} details the construction of the novel parameterized extended correlation tensor for bipartite states and presents the corresponding separability criterion, followed by two examples for detecting entanglement. Sect.~\ref{section: multipartite} generalizes the parameterized extended correlation tensor to $N$-partite systems using the generalized matricization. Biseparability and full separability criteria are provided and demonstrated with examples. Sect.~\ref{section: conclusion} discusses the conclusion and outlook.

\section{Preliminaries}\label{section: preliminaries}
Let us begin by recalling some notations and definitions. For integers $M,N$ with $M\leq N$, we denote $[M,N]=\left\{M,M+1,\cdots,N\right\}$, $[N]=\left\{1,\cdots,N\right\}$ for $N\geq1$, and $[0]=\emptyset$. 

Denote $\bm{e}_{i}\in\mathbb{R}^{d}$ as the $i$-th standard unit vector of size $d$. Let $\mathbf{I}_{N}$ denote the $N\times N$ identity matrix. 
For matrices $\mathbf{A},\mathbf{B}\in\mathbb{C}^{M\times N}$, their Hilbert-Schmidt inner product is defined as $\langle \mathbf{A},\mathbf{B}\rangle = {\rm Tr}\left(\mathbf{A}^{\dagger}\mathbf{B}\right)$; the trace norm of $\mathbf{A}$ is defined as $\left\|\mathbf{A}\right\|_{\rm tr}={\rm Tr}\left(\sqrt{\mathbf{A}^{\dagger}\mathbf{A}}\right)$, where ${\rm Tr}\left(\mathbf{A}\right)$ is the trace of $\mathbf{A}$ and $\mathbf{A}^{\dagger}$ is the transpose conjugate of $\mathbf{A}$.

A tensor is a multi-dimensional array. For tensors $\mathcal{A}=\left(a_{i_1\cdots i_N}\right)\in\mathbb{C}^{d_1\times \cdots \times d_N}$, $\mathcal{B}=\left(b_{j_1\cdots j_M}\right)\in\mathbb{C}^{d_1'\times\cdots\times d_M'}$, and a matrix $\mathbf{U}=\left(u_{i_k'i_k}\right)\in\mathbb{C}^{d_k'\times d_k}$, where $M\leq N$ and $k\in[N]$, the Frobenius norm of $\mathcal{A}$ is defined as $\left\|\mathcal{A}\right\|_F= \sqrt{\sum_{i_1,\cdots,i_N=1}^{d_1,\cdots,d_N}\left|a_{i_1\cdots i_N}\right|^2}$;
the tensor product of $\mathcal{A},\mathcal{B}$ is defined as $\mathcal{A}\otimes\mathcal{B} = \left(c_{k_1\cdots k_{N}}\right)$
with size $(d_1d_1')\times\cdots\times (d_{M}d_{M}')\times d_{M+1}\times\cdots\times d_{N}$, where $c_{k_1\cdots k_{N}} = a_{i_1\cdots i_N}b_{j_1\cdots j_M}$ and
\begin{equation*}
	k_l =\begin{cases}
		(i_l-1) d_l + j_l, & \mbox{if } l\in[M], \\
		i_l, & \mbox{if } l\in[M+1,N];
	\end{cases}
\end{equation*}
the outer product of $\mathcal{A},\mathcal{B}$ is defined as $\mathcal{A}\circ\mathcal{B} = \mathcal{C} = \left(c_{i_1\cdots i_Nj_1\cdots j_M}\right)$ with size $d_1\times\cdots\times d_N\times d_1'\times\cdots\times d_M'$, where $c_{i_1\cdots i_Nj_1\cdots j_M}=a_{i_1\cdots i_N}b_{j_1\cdots j_M}$; 
the $k$-mode product of $\mathcal{A}$ and $\mathbf{U}$ is denoted as $\mathcal{A}\times_k\mathbf{U}$ with size $d_1\times\cdots\times d_{k-1}\times d_k' \times d_{k+1}\times \cdots \times d_N$, whose entries are given by $\left(\mathcal{A}\times_k\mathbf{U}\right)_{i_1\cdots i_{k-1}i_{k}'i_{k+1}\cdots i_N} = \sum_{i_k=1}^{d_k} a_{i_1\cdots i_{k-1}i_ki_{k+1}\cdots i_N}u_{i_k'i_k}$.

For $k\in[N]$, the $k$-mode matrix unfolding~\cite{delathauwer2000multilinear} of a tensor $\mathcal{A}=\sum_{i_1,\cdots,i_N=1}^{d_1,\cdots,d_N}a_{i_1\cdots i_N}\bm{e}_{i_1}^{(1)}\circ\cdots\circ\bm{e}_{i_N}^{(N)}$ is a matrix $\mathbf{A}_{(k)}$ of size $d_k\times  \left(\prod_{j\in[N]\setminus\{k\}}d_j\right)$, defined by
\begin{equation*}
    \begin{aligned}
        \mathbf{A}_{(k)} = \sum_{i_1,\cdots,i_N=1}^{d_1,\cdots,d_N} a_{i_1\cdots i_N} \bm{e}_{i_1}^{(1)} \left(\left(\bigotimes_{j=k+1}^{N}\bm{e}_{i_j}^{(j)}\right)\otimes\left(\bigotimes_{j=1}^{k-1}\bm{e}_{i_j}^{(j)}\right)\right)^{\rm T}.
    \end{aligned}
\end{equation*}
The $k$-mode matrix unfolding of tensors maps only one dimension to the rows of the resulting matrix and the others to the columns. Naturally, one can map multiple dimensions to the rows, leading to the generalized matricization~\cite{brett2006algorithm,kolda2009tensor}. 
Let $R=\left\{r_1,\cdots,r_k\right\}$ and $C=\left\{c_1,\cdots,c_{N-k}\right\}$ be two disjoint ascending ordered sets referred to as the rows and the columns, where $R\cup C=[N]$ and $k\in[0,N]$. The unfolding is performed in backward cyclic order with two specified modes, similar to Ref.~\cite{delathauwer2000multilinear}. For $n\in[k]$ and $m\in[N-k]$, we refer to the ${(R,n;C,m)}$-mode matrix unfolding of $\mathcal{A}$ as the generalized matricization, which gives a matrix of size $\left(\prod_{i\in R} d_i\right)\times\left(\prod_{i\in C}d_i\right)$, defined by   
\begin{equation*}
    \begin{aligned}
        \mathbf{A}_{(R,n;C,m)} = & \sum_{i_1,\cdots,i_N=1}^{d_1,\cdots,d_N}a_{i_1\cdots i_N} \left(\left(\bigotimes_{j=n+1}^{k}\bm{e}_{i_{r_j}}^{(r_j)}\right)\otimes\left(\bigotimes_{j=1}^{n}\bm{e}_{i_{r_j}}^{(r_j)}\right)\right) \\
        & \cdot \left(\left(\bigotimes_{j=m+1}^{N-k}\bm{e}_{i_{c_j}}^{(c_j)}\right)\otimes\left(\bigotimes_{j=1}^{m}\bm{e}_{i_{c_j}}^{(c_j)}\right)\right)^{\rm T}.
    \end{aligned}
\end{equation*}
For a tensor $\mathcal{A}\in\mathbb{C}^{d_1\times\cdots\times d_N}$, its trace norm can be defined as $\left\|\mathcal{A}\right\|_{\rm tr} = \max_{\substack{R\cup C=[N], R\cap C=\emptyset\\ \left|R\right|=k>0,\left|C\right|=N-k>0\\ n\in[k],m\in[N-k]}}\left\{\left\|\mathbf{A}_{(R,n;C,m)}\right\|_{\rm tr}\right\}$.

The vectorization of a matrix $\mathbf{A}=\left(a_{ij}\right)$ is a vector ${\rm vec}\left(\mathbf{A}\right) = \left(a_{11}, \cdots, a_{M 1}, a_{12}, \cdots, a_{M2}, \cdots\!, a_{1N}, \cdots,a_{MN}\right)^{\rm T}$, where the element $a_{ij}$ is at the position with the row index $\left(j-1\right)M + i$. The $k$-mode vectorization of a tensor $\mathcal{A}$ is defined by ${\rm vec}_k\left(\mathcal{A}\right) = {\rm vec}\left(\mathbf{A}_{(k)}\right)$. Especially, for a vector $\bm{v}$ and a matrix $\mathbf{A}$, it holds that ${\rm vec}_1\left(\bm{v}\right)=\bm{v}$, ${\rm vec}_1\left(\mathbf{A}\right)={\rm vec}\left(\mathbf{A}\right)$, and ${\rm vec}_2\left(\mathbf{A}\right)={\rm vec}\left(\mathbf{A}^{\rm T}\right)$.

\section{Separability criterion for bipartite systems}\label{section: bipartite}
In this section, based on the generalized Bloch representation, we present a novel construction of the extended correlation tensor for bipartite quantum states, and derive the corresponding separability criterion. Two examples are provided to demonstrate the enhanced entanglement detection ability of the criterion. Moreover, we explore the relations between our criterion and the existing ones. 

\subsection{Separability criterion}
Consider a $d$-dimensional Hilbert space $\mathcal{H}_d$. Let $\mathcal{L}\left(\mathcal{H}_d\right)$ be the space of linear operators acting on $\mathcal{H}_d$ and $\left\{\mathbf{G}_i\right\}_{i=0}^{d^2-1}$ be an orthogonal basis in $\mathcal{L}\left(\mathcal{H}_d\right)$ satisfying
$\mathbf{G}_0=\mathbf{I}_d$, $\langle \mathbf{G}_i,\mathbf{G}_j\rangle = \kappa \delta_{ij}$ and ${\rm Tr}(\mathbf{G}_i)=0$, where $\delta_{ij}$ is the Kronecker delta function, $\kappa\geq1$ is the normalized constant of the basis, $i,j\in[d^2-1]$. The value of $\kappa$ depends on the detailed choice of orthogonal basis, for example, $\kappa=2$ 
($\kappa=d$) if the orthogonal basis is chosen to be the Pauli operators (Weyl operators or Heisenberg-Weyl operators).

A quantum state $\rho\in \mathcal{H}_d$ can be expanded in the basis $\left\{\mathbf{G}_i\right\}_{i=0}^{d^2-1}$, yielding the following generalized Bloch representation,
\begin{equation}\label{generalized bloch representation}
	\rho=\frac{1}{d}\left(\mathbf{I}_d+\sum_{i=1}^{d^2-1}t_i\mathbf{G}_i\right),
\end{equation}
where the coefficient $t_i=\frac{d}{\kappa}{\rm Tr}\left(\rho \mathbf{G}_i\right)$. $\mathcal{T}=\left(t_i\right)\in\mathbb{C}^{d^2-1}$ is the correlation tensor (or generalized Bloch vector) of $\rho$.
Ref.~\cite{huang2024unifying} upper bounded the Frobenius norm of $\mathcal{T}$.
\begin{lemma}\cite{huang2024unifying}\label{lemma: bloch vector upper bound}
	For any quantum state $\rho\in\mathcal{H}_{d}$, its correlation tensor $\mathcal{T}$ satisfies $\left\|\mathcal{T}\right\|_F^2\leq\frac{d^2-d}{\kappa}$.
\end{lemma}

Consider a bipartite quantum system $\mathcal{H}_{d_A} \otimes \mathcal{H}_{d_B}$ with dimensions $d_A$ and $d_B$ of subsystems , respectively. Choose arbitrary orthogonal bases  
\begin{equation}\label{equation: base bipartite}
	\left\{\mathbf{G}_{i}^{(A)}\right\}_{i=0}^{d_A^2-1},\quad
	\left\{\mathbf{G}_{j}^{(B)}\right\}_{j=0}^{d_B^2-1},
\end{equation}
in the operator spaces $\mathcal{L}(\mathcal{H}_{d_A} )$ and $\mathcal{L}(\mathcal{H}_{d_B} )$, respectively, satisfying
\begin{equation}\label{equation: base condition bipartite}
	\begin{gathered}
		\mathbf{G}^{(A)}_0=\mathbf{I}^{(A)}_{d_A}, \langle \mathbf{G}_{i}^{(A)}, \mathbf{G}_{i'}^{(A)}\rangle=\kappa_A \delta_{ii'}, {\rm Tr}\left(\mathbf{G}_{i}^{(A)}\right) = 0, \\
		\mathbf{G}^{(B)}_0=\mathbf{I}^{(B)}_{d_B},  \langle \mathbf{G}_{j}^{(B)}, \mathbf{G}_{j'}^{(B)}\rangle=\kappa_B \delta_{jj'}, {\rm Tr}\left(\mathbf{G}_{j}^{(B)}\right) = 0, 
	\end{gathered}
\end{equation}
where $\kappa_A,\kappa_B\geq1$, $i,i'\in[d_A^2-1]$, $j,j'\in[d_B^2-1]$.
Any bipartite state $\rho^{(AB)}\in \mathcal{H}_{d_A} \otimes \mathcal{H}_{d_B}$ can be expanded in the orthogonal bases specified by Eq.~\eqref{equation: base bipartite}, yielding the following generalized Bloch representation,
\begin{equation}\label{generalized bloch representation bipartite}
	\begin{aligned}
		\rho^{(AB)} =& \frac{1}{d_A d_B}\left(\mathbf{I}_{d_A}^{(A)}\otimes \mathbf{I}_{d_B}^{(B)} \right. \\
		& + \sum_{i=1}^{d_A^2-1} t_i^{(A)} \mathbf{G}_i^{(A)} \otimes \mathbf{I}_{d_B}^{(B)} + \sum_{j=1}^{d_B^2-1} t_j^{(B)}\mathbf{I}_{d_A}^{(A)}\otimes \mathbf{G}_j^{(B)} \\
		& \left. + \sum_{i=1}^{d_A^2-1}\sum_{j=1}^{d_B^2-1}t_{ij}^{(AB)}\mathbf{G}_i^{(A)}\otimes \mathbf{G}_j^{(B)}\right),
	\end{aligned}
\end{equation}
where the coefficients are given by
\begin{equation*}
	\begin{gathered}
		t_i^{(A)}=\frac{d_A}{\kappa_A}\mathrm{Tr}\left( \rho^{(AB)} \left(\mathbf{G}_i^{(A)}\otimes \mathbf{I}_{d_B}^{(B)}\right)\right),\\
		t_j^{(B)}=\frac{d_B}{\kappa_B}\mathrm{Tr}\left( \rho^{(AB)} \left(\mathbf{I}_{d_A}^{(A)}\otimes \mathbf{G}_j^{(B)}\right)\right),\\
		t_{ij}^{(AB)}=\frac{d_A d_B}{\kappa_A \kappa_B}\mathrm{Tr}\left( \rho^{(AB)} \left(\mathbf{G}_i^{(A)}\otimes \mathbf{G}_j^{(B)}\right)\right).
	\end{gathered}
\end{equation*}
$\mathcal{T}^{(A)} = \left(t_i^{(A)}\right)\in\mathbb{C}^{d_A^2-1}$, $\mathcal{T}^{(B)} = \left(t_j^{(B)}\right)\in\mathbb{C}^{d_B^2-1}$, and $\mathcal{T}^{(AB)} =(t_{ij}^{(AB)})\in\mathbb{C}^{(d_A^2-1)\times (d_B^2-1)}$ are the correlation tensors of $\rho^{(AB)}$ for the corresponding (sub)systems. 

Given a quantum state $\rho^{(AB)}\in\mathcal{H}_{d_A} \otimes \mathcal{H}_{d_B}$, we construct the extended correlation tensor based on the correlation tensor and the reduced density matrices. For any vectors $\bm{u}=\left(u_1,\cdots,u_{p_1}\right)^{\rm T}\in\mathbb{R}^{p_1}$, $\bm{v}=\left(v_1,\cdots,v_{p_2}\right)^{\rm T}\in\mathbb{R}^{P_2}$, $\bm{\alpha}=\left(a_1, \cdots, a_{q_1}\right)^{\rm T}\in\mathbb{R}^{q_1}$, and $\bm{\beta}=\left(b_1, \cdots, b_{q_2}\right)^{\rm T}\in\mathbb{R}^{q_2}$, the extended correlation tensor has a size of $\left(p_1+q_1\left(d_A^2-1\right)\right)\times\left(p_2+q_2\left(d_B^2-1\right)\right)$ and is defined by
\begin{equation}\label{extended correlation tensor bipartite}
	\mathcal{M}_{\bm{u},\bm{v},\bm{\alpha},\bm{\beta}}^{(A|B)}(\rho^{(AB)}) = \begin{pmatrix}
		\bm{u}\bm{v}^{\rm T} & \bm{u}\left(\bm{\beta}\otimes \mathcal{T}^{(B)}\right)^{\rm T} \\
		\left(\bm{\alpha} \otimes \mathcal{T}^{(A)}\right)\bm{v}^{\rm T} & \bm{\alpha}\bm{\beta}^{\rm T} \otimes \mathcal{T}^{(AB)}\\
	\end{pmatrix}.
\end{equation}

A separable state $\rho^{(AB)}$ can be expressed as a convex combination of product states, 
$\rho^{(AB)} = \sum_i p_i \rho^{(A)}_{i} \otimes \rho^{(B)}_{i}$, where $0\leq p_i\leq1$, $\sum_{i}p_i=1$, $\rho^{(A)}_{i}$ and $\rho^{(B)}_{i}$ are pure states in $\mathcal{H}_{d_A}$ and $\mathcal{H}_{d_B}$, respectively. 
Based on the extended correlation tensor, we present the main theorem of bipartite separability criterion, see proof in Appendix~\ref{section: proof of theorem 1}.

\begin{theorem}\label{theorem: bipartite separability criteria}
	If a quantum state $\rho^{(AB)}\in\mathcal{H}_{d_A} \otimes \mathcal{H}_{d_B}$ is separable, then its extended correlation tensor given by Eq.~\eqref{extended correlation tensor bipartite} satisfies
	\begin{equation}\label{Thm1}
		\begin{aligned}
			& \left\|  \mathcal{M}_{\bm{u},\bm{v},\bm{\alpha},\bm{\beta}}^{(A|B)} (\rho^{(AB)}) \right\|_{\rm tr} \\
			\leq& \sqrt{\left(\left\|\bm{u}\right\|_F^2+\left\|\bm{\alpha}\right\|_F^2\frac{{d_A}^2-d_A}{ \kappa_A}\right) \left(\left\|\bm{v}\right\|_F^2+\left\|\bm{\beta}\right\|_F^2\frac{{d_B}^2-d_B}{ \kappa_B}\right)}.
		\end{aligned}
	\end{equation}
\end{theorem}

Using the bipartite separability criterion given in Theorem~\ref{theorem: bipartite separability criteria}, we detect the entanglement of two bipartite quantum states in the following examples.

\begin{example}
	\label{example: bipartite ppt state}
	Consider the $3 \times 3$ PPT entangled state~\cite{bennett1999unextendible},
	\begin{equation*}
		\rho = \frac{1}{4}\left(\mathbf{I}_9 - \sum_{i=0}^{4}\ket{\varphi_i}\bra{\varphi_i}\right),
	\end{equation*}
	where $\ket{\varphi_0}=\frac{1}{\sqrt{2}}\ket{0}\left(\ket{0}-\ket{1}\right)$, $\ket{\varphi_1}=\frac{1}{\sqrt{2}}\left(\ket{0}-\ket{1}\right)\ket{2}$, $\ket{\varphi_2}=\frac{1}{\sqrt{2}}\ket{2}\left(\ket{1}-\ket{2}\right)$, $\ket{\varphi_3}=\frac{1}{\sqrt{2}}\left(\ket{1}-\ket{2}\right)\ket{0}$, and $\ket{\varphi_4}=\frac{1}{3}\left(\ket{0}+\ket{1}+\ket{2}\right)\left(\ket{0}+\ket{1}+\ket{2}\right)$. We mix $\rho$ with white noise and consider the following state, 
	\begin{equation}\label{equation: state example 1}
		\rho_p=\frac{1-p}{9}\mathbf{I}_9 + p\rho,
	\end{equation}
	where $0\leq p\leq1$. 
	in
	Choose the Heisenberg-Weyl operators as orthogonal bases of the bipartite system $\mathcal{H}_{d_A}\otimes\mathcal{H}_{d_B}$, with $\kappa_A=d_A=\kappa_B=d_B=3$. Taking $\bm{u}=\left(x,x,x,x\right)^{\rm T}$, $\bm{v}=\left(\sqrt{x},\sqrt{x},\sqrt{x},\sqrt{x}\right)^{\rm T}$, $\bm{\alpha}=0.063782$ and $\bm{\beta}=0.0786454$, by Theorem~\ref{theorem: bipartite separability criteria} we detect that $\rho_p$ is entangled for $0.882252\leq p\leq1$. 
	The result is shown in Figure~\ref{fig: example 1}, where the blue plane surface represents the zero plane, and the yellow curved surface represents the difference between the trace norm and the upper bound   on the left and right hand sides of Eq.~\eqref{Thm1}. Existing separability criteria such as CCNR~\cite{rudolph2005further}, Li’s criterion~\cite{li2014quantum}, dV’s criterion~\cite{devicente2007separability}, and Huang's criterion~\cite{huang2024unifying} can also detect the entanglement of the state $\rho_p$.
	\begin{figure}[h!]
		\centering
		\includegraphics[width=0.5\textwidth]{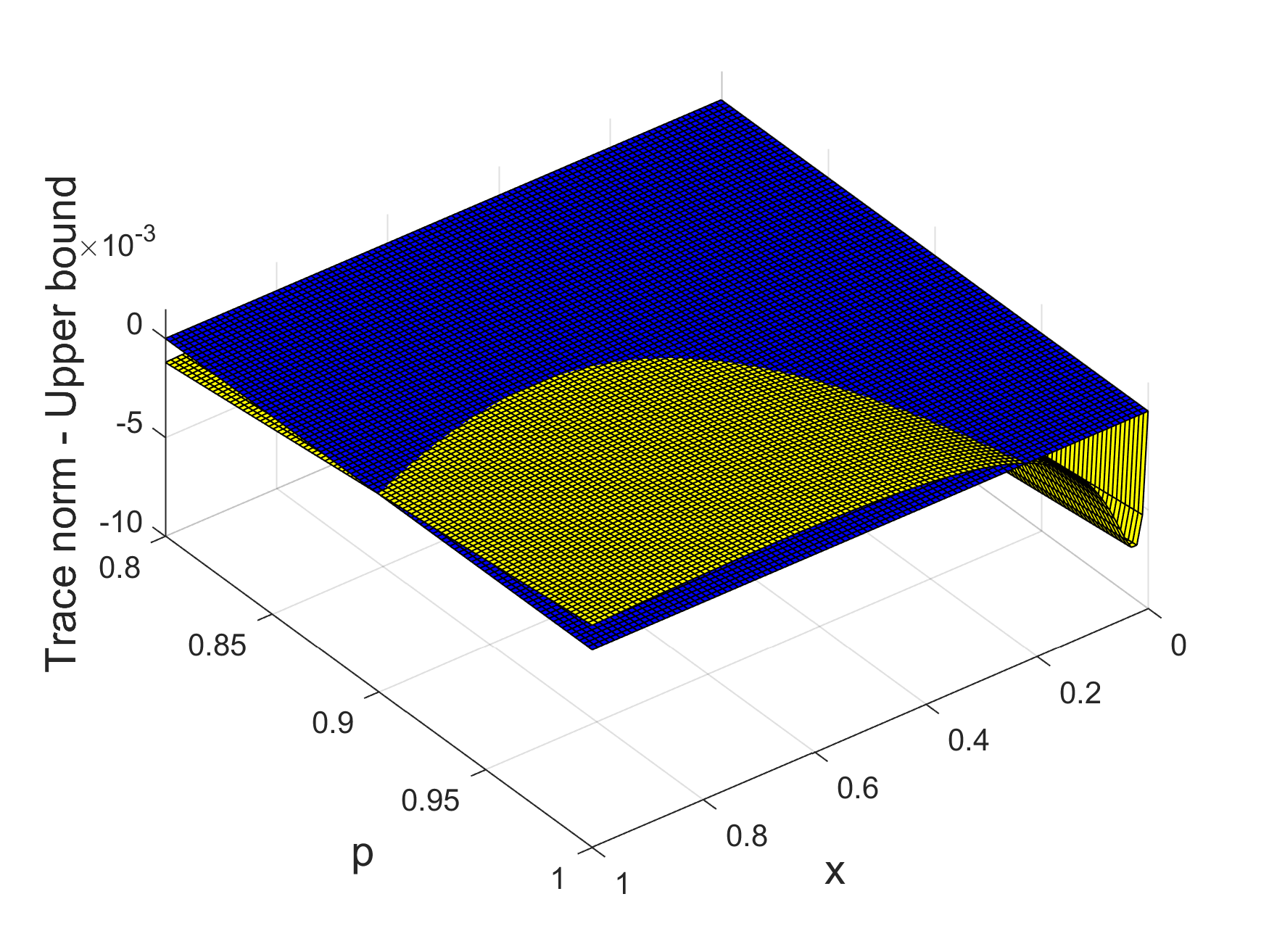}
		\caption{The relation between the entanglement $\rho_p$ and the parameters $p,x$ in Example~\ref{example: bipartite ppt state}, with $\bm{u}=\left(x,x,x,x\right)^{\rm T}$, $\bm{v}=\left(\sqrt{x},\sqrt{x},\sqrt{x},\sqrt{x}\right)^{\rm T}$, $\bm{\alpha}=0.063782$ and $\bm{\beta}=0.0786454$.}
		\label{fig: example 1}
	\end{figure}
	
	Table~\ref{table: example 1} shows the comparison between their detection thresholds and ours. It is observed that our Theorem~\ref{theorem: bipartite separability criteria} detects the entanglement of the state $\rho_p$ in Eq.~\eqref{equation: state example 1} better.
	\begin{table}[h!]
		\centering
		\caption{The comparison of detection thresholds for various separability criteria for the state $\rho_p$ in Eq.~\eqref{equation: state example 1}.}
		\begin{tabular}{>{\centering\arraybackslash}m{3cm}>{\centering\arraybackslash}m{4cm}}
			\hline
			\hline
			\textbf{Separability criteria} & \textbf{Range (entanglement) of $p$} \\
			\hline
			Theorem~\ref{theorem: bipartite separability criteria} & $0.882252\leq p\leq1$ \\
			Huang's criterion~\cite{huang2024unifying} & $0.8843<p\leq1$ \\
			Li's criterion~\cite{li2014quantum} & $0.8925<p\leq1$ \\
			CCNR~\cite{rudolph2005further} & $0.8897<p\leq1$ \\
			dV's criterion~\cite{devicente2007separability} & $0.9493<p\leq1$ \\
			\hline
			\hline
		\end{tabular}
		\label{table: example 1}
	\end{table}
\end{example}

\begin{example}\label{example: bipartite bounded entangled state}
We consider the state
\begin{equation}\label{equation: state example 2}
	\rho_p = p\ket{\xi}\bra{\xi} + (1-p)\rho_a,
\end{equation}
where $0\leq p\leq 1$,
\begin{equation*}
	\rho_a = \frac{1}{7a+1}\begin{pmatrix}
		a & 0 & 0 & 0 & 0 & a & 0 & 0 \\
		0 & a & 0 & 0 & 0 & 0 & a & 0 \\
		0 & 0 & a & 0 & 0 & 0 & 0 & a \\
		0 & 0 & 0 & a & 0 & 0 & 0 & 0 \\
		0 & 0 & 0 & 0 & \frac{1+a}{2} & 0 & 0 & \frac{\sqrt{1-a^2}}{2} \\
		a & 0 & 0 & 0 & 0 & a & 0 & 0 \\
		0 & a & 0 & 0 & 0 & 0 & a & 0 \\
		0 & 0 & a & 0 & \frac{\sqrt{1-a^2}}{2} & 0 & 0 & \frac{1+a}{2}
	\end{pmatrix}
\end{equation*}
is the $2\times4$ bounded entangled state~\cite{horodechi1997separability} and
$\ket{\xi} = \frac{1}{\sqrt{2}}\left(\ket{00}+\ket{11}\right)$, $0<a<1$. 

Choose the Pauli operators as the orthogonal bases of the system. So we have $\kappa_A=\kappa_B=2$. Set $a=0.9$, $\bm{u}=\left(x,x\right)^{\rm T}$, $\bm{v}=\left(\sqrt{x},\sqrt{x}\right)^{\rm T}$, $\bm{\alpha}=\left(-3.23405, 1.35293\right)^{\rm T}$ and $\bm{\beta}=\left(-1.83346, -0.969888\right)^{\rm T}$. By Theorem~\ref{theorem: bipartite separability criteria}, the state $\rho_p$ is entangled for all $0\leq p\leq1$. The result is shown in Figure~\ref{fig: example 2}, where the blue plane surface represents zero plane, and the yellow curved surface represents the difference between the trace norm and the upper bound  on the left and right hand sides of Eq.~\eqref{Thm1}. Compared with the existing separability criteria, our Theorem~\ref{theorem: bipartite separability criteria} clearly detects better the entanglement of the state $\rho_p$ in Eq.~\eqref{equation: state example 2}, see Table \ref{table: example 2}.
\begin{figure}[h!]
	\centering
	\includegraphics[width=0.5\textwidth]{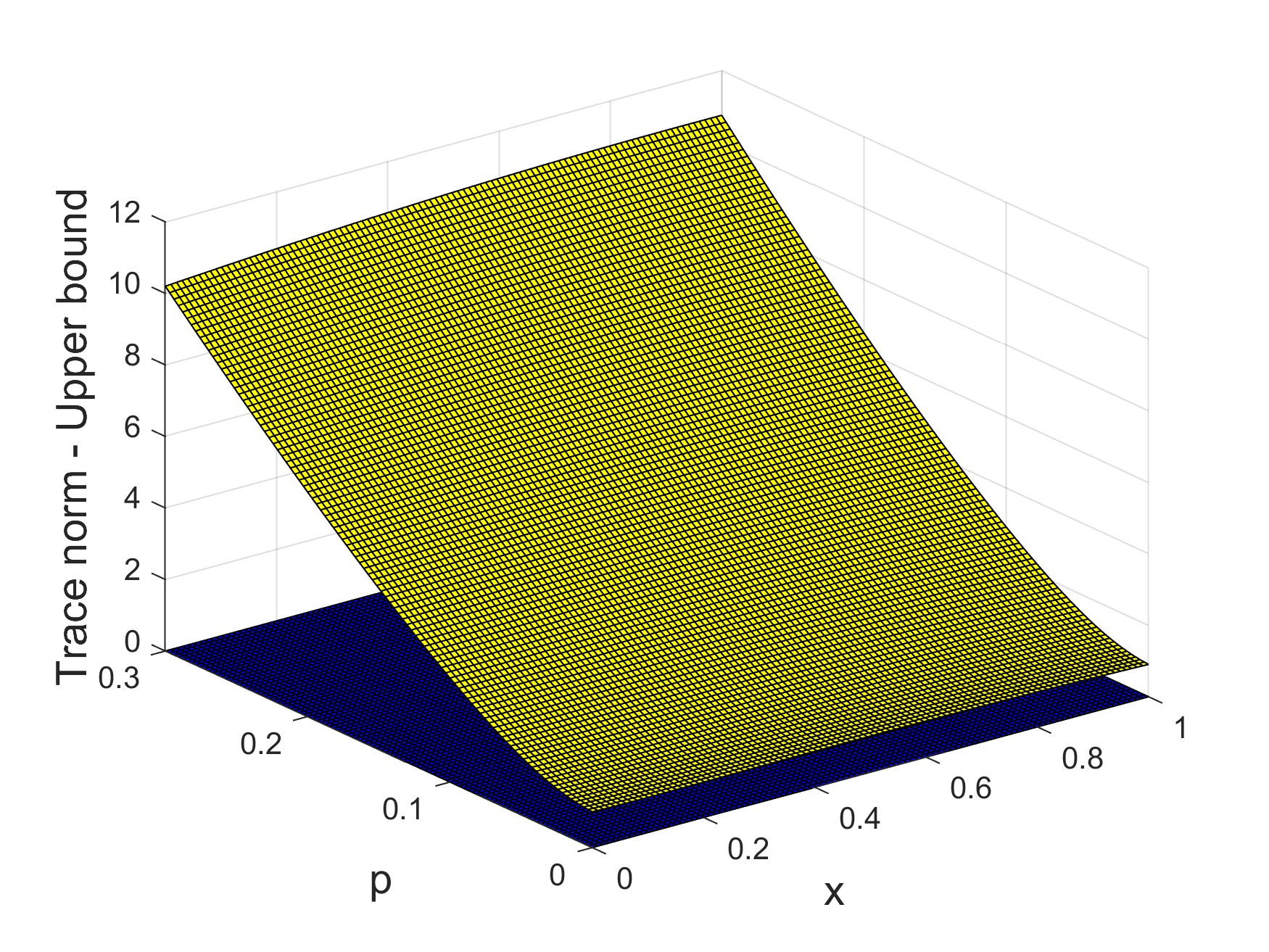}
	\caption{The relation between the entanglement of $\rho_p$ and the parameters $p,x$ in Example~\ref{example: bipartite bounded entangled state} with $a=0.9$, $\bm{u}=\left(x,x\right)^{\rm T}$, $\bm{v}=\left(\sqrt{x},\sqrt{x}\right)^{\rm T}$, $\bm{\alpha}=\left(-3.23405, 1.35293\right)^{\rm T}$ and $\bm{\beta}=\left(-1.83346, -0.969888\right)^{\rm T}$.}
	\label{fig: example 2}
\end{figure}
\begin{table}[h!]
	\centering
	\caption{The comparison of detection thresholds for various separability criteria for the state $\rho_p$ in Eq.~\eqref{equation: state example 2} with $a=0.9$.}
	\begin{tabular}{>{\centering\arraybackslash}m{3cm}>{\centering\arraybackslash}m{4cm}}
		\hline
		\hline
		\textbf{Separability criteria} & \textbf{Range (entanglement) of $p$} \\
		\hline
		Theorem~\ref{theorem: bipartite separability criteria} & $0\leq p\leq1$ \\
		Sun's criterion~\cite{sun2025separability} & $0.233889<p\leq1$ \\
		Shen's criterion~\cite{shen2016improved} & $0.2235<p\leq1$ \\
		dV's criterion~\cite{devicente2007separability} & $0.2293<p\leq1$ \\
		Li's criterion~\cite{li2014quantum} & $0.2841<p\leq1$ \\
		\hline
		\hline
	\end{tabular}
	\label{table: example 2}
\end{table}    
\end{example}

\subsection{Relations to other separability criteria}
The separability criterion for bipartite quantum systems proposed in this paper introduces parameters $\bm{u}=\left(u_1,\cdots,u_{p_1}\right)^{\rm T}\in\mathbb{R}^{p_1}$, $\bm{v}=\left(v_1,\cdots,v_{p_2}\right)^{\rm T}\in\mathbb{R}^{p_2}$, $\bm{\alpha}=\left(a_1, \cdots, a_{q_1}\right)^{\rm T}\in\mathbb{R}^{q_1}$ and $\bm{\beta}=\left(b_1, \cdots, b_{q_2}\right)^{\rm T}\in\mathbb{R}^{q_2}$, where $p_1,p_2,q_1,q_2$ are positive integers. It selects orthogonal bases $\left\{\mathbf{G}_{i}^{(A)}\right\}_{i=0}^{d_A^2-1}$ and $\left\{\mathbf{G}_{i}^{(B)}\right\}_{i=0}^{d_B^2-1}$ in the operator spaces $\mathcal{L}(\mathcal{H}_{d_A} )$ and $\mathcal{L}(\mathcal{H}_{d_B} )$, respectively, satisfying Eq.~\eqref{equation: base condition bipartite} with $\kappa_A,\kappa_B\geq1$. The inclusion of these parameters and the freedom in choosing orthogonal bases establish close connections between the separability criterion presented in this paper and several existing ones, while also improve the ability of entanglement detection.

In 2007, de Vicente~\cite{devicente2007separability} provided an upper bound for the trace norm of the correlation tensor $\mathcal{T}^{(AB)}$ for separable states,
\begin{equation*}
	\left\|\mathcal{T}^{(AB)}\right\|_{\rm tr} \leq \sqrt{\frac{{d_Ad_B}\left(d_A-1\right)\left(d_B-1\right)}{4}},
\end{equation*} 
where the orthogonal bases are chosen as the Pauli operators, that is, $\kappa_A=\kappa_B=2$. Taking $\bm{u}=\bm{v}=0$, $\bm{\alpha}=\bm{\beta}=1$ in Eq.~\eqref{extended correlation tensor bipartite}, we obtain
\begin{equation*}
	\begin{aligned}
		\left\|\mathcal{T}^{(AB)} \right\|_{\rm tr} =& \left\|  \mathcal{M}_{0,0,1,1}^{(A|B)} \left(\rho^{(AB)}\right) \right\|_{\rm tr} \\
		\leq& \sqrt{\left(\frac{d^2_A-d_A}{2}\right)\left(\frac{d^2_B-d_B}{2}\right)} \\
		=& \sqrt{\frac{d_Ad_B\left(d_A-1\right)\left(d_B-1\right)}{4}}.
	\end{aligned}
\end{equation*} 
which implies that the conclusion in Theorem~\ref{theorem: bipartite separability criteria} reduces to that given in Ref.~\cite{devicente2007separability} with particularly fixed parameters.

In 2016, Shen et al.~\cite{shen2016improved} introduced parameters $x,y\in\mathbb{R}$ and the matrix $\mathbf{E}_{l\times l}$ of size $l\times l$ with all  entries being one and constructed the following extended correlation tensor,
\begin{equation}\label{shen's extended correlation tensor}
	\mathcal{S}_{x,y}^{l}(\rho^{(AB)}) = \begin{pmatrix}
		xy\mathbf{E}_{l\times l} & x\omega_l^{\rm T}\left(\mathcal{T}^{(B)}\right) \\
		y\omega_l\left(\mathcal{T}^{(A)}\right) & \mathcal{T}^{(AB)} \\
	\end{pmatrix},
\end{equation}
where $\omega_l\left(\bm{v}\right)=\underbrace{\left(\bm{v},\cdots,\bm{v}\right)}_{l}$ and the orthogonal bases are chosen as the Pauli operators. It has been proven that for the separable quantum state $\rho^{(AB)}$,
\begin{equation*}
	\begin{aligned}
		\left\|\mathcal{S}_{x,y}^{l}(\rho^{(AB)})  \right\|_{\rm tr} &\leq \frac{1}{2}\sqrt{\left(2lx^2+d^2_A-d_A\right)\left(2ly^2+d^2_B-d_B\right)} .
	\end{aligned}
\end{equation*} 
Using the extended correlation tensor given in Eq.~\eqref{extended correlation tensor bipartite} and taking $\bm{u}=\left(x,\cdots,x\right)^{\rm T}\in\mathbb{R}^{l}$, $\bm{v}=\left(y,\cdots,y\right)^{\rm T}\in\mathbb{R}^{l}$, $\bm{\alpha}=\bm{\beta}=1$, we have
\begin{equation*}
	\begin{aligned}
		& \left\|\mathcal{S}_{x,y}^{l}(\rho^{(AB)})  \right\|_{\rm tr} =\left\|  \mathcal{M}_{\bm{u},\bm{v},1,1}^{(A|B)} (\rho^{(AB)}) \right\|_{\rm tr} \\
		\leq& \sqrt{\left(lx^2+\frac{d^2_A-d_A}{2}\right)\left(ly^2+\frac{d^2_B-d_B}{2}\right)}\\
		=& \frac{1}{2}\sqrt{\left(2lx^2+d^2_A-d_A\right)\left(2ly^2+d^2_B-d_B\right)}.
	\end{aligned}
\end{equation*} 
It can be seen that our separability criterion covers the one given in Ref.~\cite{shen2016improved}. In addition, if we take $\kappa_A,\kappa_B>2$, then the trace norm of our extended correlation tensor in Eq.~\eqref{extended correlation tensor bipartite} has a lower upper bound.

In 2018, Chang et al.~\cite{chang2018separability} constructed the extended correlation tensor that is structurally identical to Eq.~\eqref{shen's extended correlation tensor}, but adopted the Heisenberg-Weyl operators as orthogonal bases, resulting in $\kappa_A=d_A$ and $\kappa_B=d_B$. Then they provided the following separable criterion,
\begin{equation*}
	\left\|\mathcal{S}_{x,y}^{l}(\rho^{(AB)}) \right\|_{\rm tr} \leq \sqrt{\left(lx^2+d_A-1\right)\left(ly^2+d_B-1\right)},
\end{equation*} 
which is also a special case of our separability criterion with $\bm{u}=\left(x,\cdots,x\right)^{\rm T}\in\mathbb{R}^{l}$, $\bm{v}=\left(y,\cdots,y\right)^{\rm T}\in\mathbb{R}^{l}$ and $\bm{\alpha}=\bm{\beta}=1$.

In 2020, Sarbicki et al.~\cite{sarbicki2020family} employed the canonical basis to establish a generalized Bloch representation for bipartite quantum states, where the canonical bases $\left\{\mathbf{G}_{i}^{(A)}\right\}_{i=0}^{d_A^2-1}$ and $\left\{\mathbf{G}_{i}^{(B)}\right\}_{i=0}^{d_B^2-1}$ are orthogonal and satisfy $\langle \mathbf{G}_{i}^{(A)}, \mathbf{G}_{j}^{(A)}\rangle=\delta_{ij}$, $\langle \mathbf{G}_{i}^{(B)}, \mathbf{G}_{j}^{(B)}\rangle=\delta_{ij}$. The resulting coefficients of generalized Bloch representation form a matrix $\mathbf{C}^{\rm can}\in\mathbb{C}^{d^2_A\times d^2_B}$, from which they derived the following separability criterion,
\begin{equation*}
	\left\|\mathbf{D}^A_x \mathbf{C}^{\rm can} \mathbf{D}^B_y \right\|_{\rm tr} \leq \sqrt{\left(\frac{d_A-1+x^2}{d_A}\right)\left(\frac{d_B-1+y^2}{d_B}\right)},
\end{equation*} 
where $\mathbf{D}^A_x={\rm diag}\left\{x,1,\dots,1\right\}\in\mathbb{R}^{d^2_A\times d^2_A}$ and $\mathbf{D}^B_y={\rm diag}\left\{y,1,\dots,1\right\}\in\mathbb{R}^{d^2_B\times d^2_B}$, $x,y\geq 0$. In fact, the matrix $\mathbf{D}^A_x \mathbf{C}^{\rm can} \mathbf{D}^B_y$ is equivalent to a special case of the extended correlation tensor proposed in this work. Taking $\bm{u}=\frac{x}{\sqrt{d_A}}$, $\bm{v}=\frac{y}{\sqrt{d_B}}$, $\bm{\alpha}=\bm{\beta}=1$, $\kappa_A=\kappa_B=1$ in Eq.~\eqref{extended correlation tensor bipartite}, we obtain
\begin{equation*}
	\begin{aligned}
		& \left\| \mathcal{M}_{\bm{u},\bm{v},1,1}^{(A|B)} \left(\rho^{(AB)}\right) \right\|_{\rm tr} \\
		\leq& \sqrt{\left(\frac{d^2_A\left(d_A-1\right)+x^2}{d_A}\right)\left(\frac{d^2_B\left(d_B-1\right)+y^2}{d_B}\right)},
	\end{aligned}
\end{equation*} 
where the upper bound differs from the one in Ref.~\cite{sarbicki2020family} by additional factors $d_A^2$ and $d_B^2$, because the generalized Bloch representations in Ref.~\cite{sarbicki2020family} and this paper differ by a scale $\frac{1}{d_Ad_B}$ so that the upper bounds of the correlation tensor for a monomer quantum state used in the proof differ by a scale $d^2$. 

In 2023, Zhu et al.~\cite{zhu2023family} introduced the following extended correlation tensor,
\begin{equation*}
	\mathcal{T}_{\bm{u}\bm{v}}\left(\rho^{(AB)}\right) = \begin{pmatrix}
		\bm{u}\bm{v}^{\rm T} & \bm{u}\left(\mathcal{T}^{(B)}\right)^{\rm T} \\
		\mathcal{T}^{(A)}\bm{v}^{\rm T} & \mathcal{T}^{(AB)}\\
	\end{pmatrix},
\end{equation*}
based on the generalized Bloch representation of quantum states, where $\bm{u}=\left(u_1,\dots,u_{p_1}\right)\in\mathbb{R}^{p_1}$, $\bm{v}=\left(v_1,\dots,v_{p_2}\right)\in\mathbb{R}^{p_2}$. From this construction, they derived the corresponding separability criterion,
\begin{equation*}
	\begin{aligned}
		& \left\|\mathcal{T}_{\bm{u}\bm{v}}\left(\rho^{(AB)}\right) \right\|_{\rm tr} \\
		\leq& \sqrt{\left(\left\|\bm{u}\right\|^2_F+\frac{d^2_A-d_A}{2}\right)\left(\left\|\bm{v}\right\|^2_F+\frac{d^2_B-d_B}{2}\right)}.
	\end{aligned}
\end{equation*} 
It can be seen that Zhu et al.'s criterion is a special case of $\bm{\alpha} = \bm{\beta} = 1$ and $\kappa_A = \kappa_B = 2$ of ours.

In 2024, Huang et al.~\cite{huang2024unifying} proposed a unifying extended correlation tensor, 
\begin{equation*}
	\mathcal{M}_{x,y}^{(n)} = \begin{pmatrix}
		\frac{xy}{\sqrt{\kappa_A\kappa_Bd_Ad_B}}\mathbf{E}_{n\times n} & \frac{x}{\sqrt{\kappa_Ad_A}}\omega_n^{\rm T}\left(\mathcal{T}^{(B)}\right) \\
		\frac{y}{\sqrt{\kappa_Bd_B}}\omega_n\left(\mathcal{T}^{(A)}\right)  & \mathcal{T}^{(AB)}
	\end{pmatrix},
\end{equation*}
for arbitrary orthogonal bases, where $x,y$ are non-negative real numbers, $n$ is a positive integer, and the generalized Bloch representation differs from ours by a scale $\frac{1}{d_Ad_B}$. From this construction, they presented the separability criterion,
\begin{equation*}
	\left\|\mathcal{M}_{x,y}^{(n)} \right\|_{\rm tr} 
	\leq \sqrt{\left(\frac{nx^2+d_A-1}{\kappa_Ad_A}\right)\left(\frac{ny^2+d_B-1}{\kappa_Bd_B}\right)}
\end{equation*}
for separable states. Setting $\bm{u}=\left(\frac{x}{\sqrt{\kappa_Ad_A}},\cdots,\frac{x}{\sqrt{\kappa_Ad_A}}\right)^{\rm T}\in\mathbb{R}^n$, $\bm{v}=\left(\frac{y}{\sqrt{\kappa_Bd_B}},\cdots,\frac{y}{\sqrt{\kappa_Bd_B}}\right)^{\rm T}\in\mathbb{R}^n$, $\bm{\alpha} = \bm{\beta} = 1$ in our extended correlation tensor $\mathcal{M}_{\bm{u},\bm{v},\bm{\alpha},\bm{\beta}}^{(A|B)}(\rho^{(AB)})$, we get
\begin{equation*}
	\begin{aligned}
		&\left\| \mathcal{M}_{\bm{u},\bm{v},1,1}^{(A|B)} \left(\rho^{(AB)}\right) \right\|_{\rm tr} \\
		\leq& \sqrt{\left(\frac{nx^2+d_A^2\left(d_A-1\right)}{\kappa_Ad_A}\right)
			\left(\frac{ny^2+d_B^2\left(d_B-1\right)}{\kappa_Bd_B}\right)},
	\end{aligned}
\end{equation*} 
where the additional factors $d_A^2$ and $d_B^2$ are owing to that the upper bounds of the correlation tensor for a monomer state differ by a scale $d^2$. 

Moreover, in Ref.~\cite{huang2024unifying} the authors also provided an extended correlation tensor by using the generalized Bloch representation with the extracted scale $\frac{1}{d_Ad_B}$,
\begin{equation*}
	\hat{\mathcal{M}}_{x,y}^{(n)} = \begin{pmatrix}
		xy\mathbf{E}_{n\times n} & x\omega_n^{\rm T}\left(\mathcal{T}^{(B)}\right) \\
		y\omega_n\left(\mathcal{T}^{(A)}\right) & \mathcal{T}^{(AB)}
	\end{pmatrix},
\end{equation*}
and the corresponding separability criterion,
\begin{equation*}
	\left\|\hat{\mathcal{M}}_{x,y}^{(n)}\right\|_{\rm tr} \leq \sqrt{\left(nx^2+\frac{d_A^2-d_A}{\kappa_A}\right) \left(ny^2+\frac{d_B^2-d_B}{\kappa_B}\right)}.
\end{equation*}
By taking $\bm{u}=\left(x,\cdots,x\right)^{\rm T}\in\mathbb{R}^n$, $\bm{v}=\left(y,\cdots,y\right)\in\mathbb{R}^n$, $\bm{\alpha}=\bm{\beta}=1$ in Eq.~\eqref{extended correlation tensor bipartite}, we have
\begin{equation*}
	\begin{aligned}
		& \left\|\hat{\mathcal{M}}_{x,y}^{(n)}\right\|_{\rm tr} = \left\| \mathcal{M}_{\bm{u},\bm{v},1,1}^{(A|B)} \left(\rho^{(AB)}\right) \right\|_{\rm tr} \\
		\leq& \sqrt{\left(nx^2+\frac{d_A^2-d_A}{\kappa_A}\right) \left(ny^2+\frac{d_B^2-d_B}{\kappa_B}\right)}.
	\end{aligned}
\end{equation*}
We have again that the criteria given in Ref.~\cite{huang2024unifying} are special cases of ours.

\section{Separability criteria for multipartite systems}\label{section: multipartite}
In this section, we construct the extended correlation tensor under the bipartition for $N$-partite quantum states using the general matrix unfolding method. The biseparability and fully separability criteria are provided with two examples to demonstrate their ability for entanglement detection.

Let $\mathcal{H}_{d_1},\mathcal{H}_{d_2},\cdots,\mathcal{H}_{d_N}$ be $N$ Hilbert spaces with dimensions $d_1,d_2,\cdots,d_N$, respectively. Denote 
\begin{equation}\label{equation: base n}
	\left\{\mathbf{G}_{i_1}^{(1)}\right\}_{i_1=0}^{d_1^2-1}, \quad
	\left\{\mathbf{G}_{i_2}^{(2)}\right\}_{i_2=0}^{d_2^2-1}, \quad
	\cdots, \quad
	\left\{\mathbf{G}_{i_N}^{(N)}\right\}_{i_N=0}^{d_N^2-1},
\end{equation}
as the orthogonal bases in the operator spaces $\mathcal{L}\left(\mathcal{H}_{d_1}\right)$, $\mathcal{L}\left(\mathcal{H}_{d_2}\right)$, $\cdots$, $\mathcal{L}\left(\mathcal{H}_{d_n}\right)$, respectively, which satisfy that
\begin{equation}\label{equation: base condition n-partite}
	\mathbf{G}^{(l)}_{0}=\mathbf{I}^{(l)}_{d_l},\quad \langle \mathbf{G}_{i_l}^{(l)}, \mathbf{G}_{i_{l}'}^{(l)} \rangle = \kappa_l \delta_{i_{l} i_{l}'},\quad
	{\rm Tr}\left(\mathbf{G}_{i_l}^{(l)}\right)=0,
\end{equation}
where $\kappa_l\geq1$, $i_l,i_{l}'\in[d_l^2-1]$, $l\in[N]$.

For an $N$-partite state $\rho \in \mathcal{H}_{d_1} \otimes \mathcal{H}_{d_2} \otimes \cdots \otimes \mathcal{H}_{d_N}$, it can be expanded in the orthogonal bases specified by Eq.~\eqref{equation: base n}, resulting the generalized Bloch representation of $\rho$,
\begin{equation}\label{generalized bloch representation n}
	\begin{aligned}
		\rho =& \frac{1}{d_1 \cdots d_N} \left( \mathbf{I}_{d_1}^{(1)}\otimes\cdots \otimes \mathbf{I}_{d_N}^{(N)} \right. \\
		& + \sum_{i_1=1}^{d_1^2-1} t_{i_1}^{(1)} \mathbf{G}_{i_1}^{(1)} \otimes \mathbf{I}_{d_2}^{(2)}\otimes\cdots\otimes \mathbf{I}_{d_N}^{(N)} +\cdots \\
		&+ \sum_{i_N=1}^{d_N^2-1} t_{i_N}^{(N)}\mathbf{I}_{d_1}^{(1)}\otimes \cdots\otimes \mathbf{I}_{d_{N-1}}^{(N-1)}\otimes \mathbf{G}_{i_N}^{(N)} \\
		& +\sum_{i_1=1}^{d_1^2-1}\sum_{i_2=1}^{d_2^2-1} t_{i_1i_2}^{(12)}\mathbf{G}_{i_1}^{(1)}\otimes \mathbf{G}_{i_2}^{(2)}\otimes \mathbf{I}_{d_3}^{(3)} \otimes\cdots\otimes \mathbf{I}_{d_N}^{(N)}  + \cdots \\
		&+ \sum_{i_{N-1}=1}^{d_{N-1}^2-1}\sum_{i_N=1}^{d_N^2-1}t_{i_{N-1}i_N}^{(N-1N)}\mathbf{I}_{d_1}^{(1)}\otimes\cdots\otimes \mathbf{I}_{d_{N-2}}^{(N-2)}\otimes \mathbf{G}_{i_{N-1}}^{(N-1)}\otimes \mathbf{G}_{i_N}^{(N)}\\
		&\left.+\cdots+\sum_{i_1=1}^{d_1^2-1}\cdots\sum_{i_N=1}^{d_N^2-1}t_{i_1\cdots i_N}^{(1\cdots N)}\mathbf{G}_{i_1}^{(1)}\otimes \cdots\otimes \mathbf{G}_{i_N}^{(N)} \right), \\
	\end{aligned}
\end{equation}
where the coefficients are
\begin{equation*}
	\begin{gathered}
		t_{i_1}^{{(1)}}=\frac{d_1}{\kappa_1}\mathrm{Tr}\left( \rho \left(\mathbf{G}_{i_1}^{(1)} \otimes \mathbf{I}_{d_2}^{(2)} \otimes\cdots\otimes \mathbf{I}_{d_N}^{(N)}\right) \right),\cdots, \\
		t_{i_N}^{{(N)}}=\frac{d_N}{\kappa_N}\mathrm{Tr}\left( \rho \left(\mathbf{I}_{d_1}^{(1)}\otimes \cdots\otimes \mathbf{I}_{d_{N-1}}^{(N-1)}\otimes \mathbf{G}_{i_N}^{(N)}\right)\right), \\
		t_{i_1i_2}^{(12)}=\frac{d_1 d_2}{\kappa_1\kappa_2}\mathrm{Tr} \left(\rho \left(\mathbf{G}_{i_1}^{(1)}\otimes \mathbf{G}_{i_2}^{(2)} \otimes \mathbf{I}_{d_3}^{(3)}\otimes\cdots\otimes \mathbf{I}_{d_N}^{(N)}\right) \right), \cdots,\\
		\begin{aligned}
			& t_{i_{N-1}i_N}^{(N-1N)} \\
			= &\frac{d_{N-1} d_N}{\kappa_{N-1}\kappa_N} \mathrm{Tr} \left(\rho \left(\mathbf{I}_{d_1}^{(1)}\otimes\cdots\otimes \mathbf{I}_{d_{N-2}}^{(N-2)}\otimes \mathbf{G}_{i_{N-1}}^{(N-1)}\otimes \mathbf{G}_{i_N}^{(N)}\right)\right),
		\end{aligned} \\
		\cdots,\\
		t_{i_1\cdots i_N}^{(1\cdots N)}=\frac{d_1d_2\cdots d_N}{\kappa_1\kappa_2\cdots \kappa_N}\mathrm{Tr} \left(\rho \left(\mathbf{G}_{i_1}^{(1)}\otimes\cdots\otimes \mathbf{G}_{i_N}^{(N)}\right) \right).
	\end{gathered}
\end{equation*}

Denote 
\begin{equation*}
	\begin{gathered}
		\mathcal{T}^{(1)}=\left(t_{i_1}^{(1)}\right)\in\mathbb{C}^{d_1^2-1}, \quad\cdots,\quad \\\mathcal{T}^{(N)}=\left(t_{i_N}^{(N)}\right)\in\mathbb{C}^{d_N^2-1}, \\
		\mathcal{T}^{(12)}=\left(t_{i_1i_2}^{(12)}\right)\in\mathbb{C}^{\left(d_1^2-1\right)\times\left(d_2^2-1\right)}, \quad\cdots,\quad \\
		\mathcal{T}^{(N-1N)}=\left(t_{i_{N-1}i_N}^{(N-1N)}\right)\in\mathbb{C}^{\left(d_{N-1}^2-1\right)\times\left(d_N^2-1\right)}, \\
		\cdots,\\
		\mathcal{T}^{(1\cdots N)} = \left(t_{i_1\cdots i_N}^{(1\cdots N)}\right)\in\mathbb{C}^{\left(d_1^2-1\right)\times\cdots\times\left(d_N^2-1\right)},
	\end{gathered}
\end{equation*}
which are the correlation tensors for the corresponding (sub)systems. 
Denote $d=\max_{i\in[N]}\left\{d_i\right\}$ and 
\begin{equation}\label{equation: m upper bound n-partite}
	\begin{aligned}
		& \mathfrak{m}^{(1\cdots N)} \\
		=& \begin{cases}
			\frac{d_1^2 - d_1}{\kappa_1}, & \mbox{if } N=1, \\
			\makecell{\tiny\begin{aligned}
				\frac{d_1 \cdots d_N}{\kappa_1\cdots\kappa_N}\left(d_1\cdots d_N + \frac{1}{N-1} \right.& \\
				\left.- \frac{d_1\cdots d_N}{N-1} \sum_{i=1}^N \frac{1}{d_i^2}\right)&,
			\end{aligned}} & \mbox{if } N=2 \mbox{ or } N\geq3 , \frac{d_1\cdots d_N}{d^2} < 1, \\
			\makecell{\tiny\begin{aligned}
				\frac{d_1 \cdots d_N}{\kappa_1\cdots\kappa_N}\left(d_1\cdots d_N + \frac{2}{N-2} \right.& \\
				\left.- \frac{d_1\cdots d_N}{N-2} \sum_{i=1}^N \frac{1}{d_i^2}\right)&,
			\end{aligned}} & \mbox{if } N\geq3, \frac{d_1\cdots d_N}{d^2} \geq 1.
		\end{cases}
	\end{aligned}
\end{equation}
The following lemma upper bounds the Frobenius norm for the correlation tensor; see proof in Appendix~\ref{section: proof of lemma 2}.

\begin{lemma}\label{lemma: upper bound of correlation tensor n-paritite}
	For a quantum state $\rho\in \mathcal{H}_{d_1}\otimes\mathcal{H}_{d_2}\otimes\cdots\otimes\mathcal{H}_{d_N}$, the correlation tensor $\mathcal{T}^{(1\cdots N)}$ of $\rho$ satisfies $\left\|\mathcal{T}^{\left(1\cdots N\right)}\right\|_F^2\leq \mathfrak{m}^{(1\cdots N)}$. 
\end{lemma}

In Eq.~\eqref{equation: m upper bound n-partite}, the second bound continues to hold for the case of $N\geq3, \frac{d_1\cdots d_N}{d^2}\geq1$, but it is not sharper than the third bound. Table~\ref{table: comparison of upper bounds} presents our results and several upper bounds in Ref.~\cite{zhao2022detection,zhang2023improved}. We compare them and find that our upper bound is either sharper than their bounds or can reduce to them, see the details in Appendix~\ref{section: upper bound of frobenius norm of correlation tensor}.
\begin{table*}[htbp]
	\centering
	\caption{Upper bounds on the Frobenius norm for the correlation tensors of $N$-partite quantum states in $\mathcal{H}_{d_1}\otimes\cdots\otimes\mathcal{H}_{d_N}$, where $d=\max_{i_[N]}\{d_i\}$.}
	\begin{tabular}{|c|c|c|c|}
		\hline
		& \multicolumn{3}{c|}{\textbf{Upper bound}} \\
		\cline{2-4}
		& $N=1$ & $N=2$ & $N\geq3$ \\
		\hline
		Lemma 3 in this paper & $\frac{d_1^2 - d_1}{\kappa_1}$ & $\frac{d_1^2d_2^2+d_1d_2 - d_1^2 - d_2^2}{\kappa_1\kappa_2}$ & $\begin{cases}
			\frac{d_1 \cdots d_N}{\kappa_1\cdots\kappa_N}\left(d_1\cdots d_N + \frac{1}{N-1}- \frac{d_1\cdots d_N}{N-1} \sum_{i=1}^N \frac{1}{d_i^2}\right), & \mbox{if } \frac{d_1\cdots d_N}{d^2} < 1 \\
			\frac{d_1 \cdots d_N}{\kappa_1\cdots\kappa_N}\left(d_1\cdots d_N + \frac{2}{N-2}- \frac{d_1\cdots d_N}{N-2} \sum_{i=1}^N \frac{1}{d_i^2}\right), & \mbox{if } \frac{d_1\cdots d_N}{d^2} \geq 1
		\end{cases}$ \\
		\hline
		\makecell{Lemma 1, 2 in Ref.~\cite{zhao2022detection} \\ (Generalized Pauli basis)} & $d_1-1$ & \multicolumn{2}{c|}{$\frac{d_1\cdots d_N\left(N-1-\sum_{i=1}^N\frac{1}{d_i^2}\right)+1}{N-1}$} \\
		\hline
		\makecell{Lemma 1, 2, 3 in Ref.~\cite{zhang2023improved} \\ (Weyl basis)} & $d_1-1$ & $d_1d_2\left(1-\frac{1}{d^2}\right)$ & \makecell{$d_1\cdots d_N - \frac{d_1\cdots d_N}{N-1}\sum_{i=1}^{N}\frac{1}{d_i^2}+\frac{1}{N-1} + \frac{1}{N-2}\left(\frac{N}{N-1} - \frac{d_1\cdots d_N}{N-1}\sum_{i=1}^{N}\frac{1}{d_i^2}\right)$ \\ if $\frac{d_1\cdots d_N}{d^2} \geq 1$} \\
		\hline
	\end{tabular}
	\label{table: comparison of upper bounds}
\end{table*}

For an $N$-partite system $\mathcal{H}_{d_1} \otimes \mathcal{H}_{d_2} \otimes \cdots \otimes \mathcal{H}_{d_N}$, denote $l_1\cdots l_k|l_{k+1}\cdots l_N$ as the bipartition of the system, where $\left\{l_1,\cdots, l_N\right\}=[N]$, $l_1<\cdots<l_k$, $l_{k+1}<\cdots<l_N$, $l_1<l_{k+1}$, $k\in[N-1]$. An $N$-partite system $\rho$ is said to be biseparable under the bipartition $l_1\cdots l_k|l_{k+1}\cdots l_N$ if it can be expressed as a convex combination of product states under the bipartition,
\begin{equation*}
	\begin{aligned}
		\rho =& \rho^{(l_l\cdots l_k|l_{k+1}\cdots l_N)} \\
		=& \sum_i p_i^{(l_1\cdots l_k|l_{k+1}\cdots l_N)} \rho_{i}^{(l_1\cdots l_k)} \otimes \rho_{i}^{(l_{k+1}\cdots l_N)},
	\end{aligned}
\end{equation*}
where $0\leq p_{i}^{(l_1\cdots l_k|l_{k+1}\cdots l_N)}\leq1$, $\sum_ip_i^{(l_1\cdots l_k|l_{k+1}\cdots l_n)}=1$,
$\rho_{i}^{(l_1\cdots l_k)}$ and $\rho_{i}^{(l_{k+1}\cdots l_N)}$ are pure states in the subsystems $\mathcal{H}_{d_{l_1}}\otimes\cdots\otimes\mathcal{H}_{d_{l_k}}$ and $\mathcal{H}_{d_{l_{k+1}}}\otimes\cdots\otimes\mathcal{H}_{d_{l_N}}$, respectively. An $N$-partite system $\rho$ is said to be biseparable if it can be expressed as a convex combination of biseparable states under bipartitions,
\begin{equation*}
	\rho = \sum_{k=1}^{N-1}\sum_{\substack{\left\{l_1,\cdots,l_N\right\}=[N]\\l_1<\cdots<l_k\\l_{k+1}<\cdots<l_N\\l_1<l_{k+1}}} p_{l_1\cdots l_N} \rho^{(l_l\cdots l_k|l_{k+1}\cdots l_N)},
\end{equation*}
where $0\leq p_{l_1\cdots l_N}\leq1$, $\sum_{k=1}^{N-1}\sum\limits_{\substack{\left\{l_1,\cdots,l_N\right\}=[N]\\l_1<\cdots<l_k\\l_{k+1}<\cdots<l_N\\l_1<l_{k+1}}} p_{l_1\cdots l_N}=1$, and $\rho^{(l_l\cdots l_k|l_{k+1}\cdots l_N)}$ is biseparable under the bipartition $l_l\cdots l_k|l_{k+1}\cdots l_N$.
Accordingly, $\rho$ is called genuine $N$-partite entangled if it is not biseparable. An $N$-partite system $\rho$ is said to be fully separable if it can be expressed as a convex combination of the tensor product of $N$ pure states, $\rho = \sum_ip_i \rho^{(1)}_i\otimes\rho^{(2)}_i\otimes\cdots\otimes\rho^{(N)}_i$,
where $\rho^{(k)}_i$ is a pure state in the subsystem $\mathcal{H}_{d_k}$, $k\in[N]$. And $\rho$ is entangled if it is not fully separable. These definitions can be found in the works of Ref.~\cite{horodechi2009quantum, hassan2008separability,laskowki2011correlation}.

We explore the property of $N$-partite biseparable states under a given bipartition in the following lemma; see proof in Appendix~\ref{section: proof of lemma 3}.

\begin{lemma}\label{lemma: property of correlation tensor n}
	Given a bipartition $l_1\cdots l_k|l_{k+1}\cdots l_N$, where $\left\{l_1,\cdots,l_N\right\}=[N]$, $l_1<\cdots<l_k$, $l_{k+1}<\cdots<l_N$, $l_1<l_{k+1}$, $k\in[N-1]$, if an $N$-partite quantum state $\rho\in\mathcal{H}_{d_1}\otimes\mathcal{H}_{d_2}\otimes\cdots\otimes\mathcal{H}_{d_N}$ is biseparable under the bipartition $l_1\cdots l_k|l_{k+1}\cdots l_N$, that is, $\rho = \sum_ip_i \rho_i^{(l_1\cdots l_k)} \otimes \rho_i^{(l_{k+1}\cdots l_N)}$,
	where $0\leq p_i\leq1$, $\sum_ip_i=1$, $\rho_i^{(l_1\cdots l_k)}$ and $\rho_i^{(l_{k+1}\cdots l_N)}$ are pure states in the subsystems $\mathcal{H}_{d_{l_1}}\otimes\cdots\otimes\mathcal{H}_{d_{l_p}}$ and $\mathcal{H}_{d_{l_{k+1}}}\otimes\cdots\otimes\mathcal{H}_{d_{l_N}}$, respectively, then 
    for any $n\in [p]$, $m\in [q]$,
	\begin{equation*}
		\mathbf{T}^{(r_1\cdots r_pc_1\cdots c_q)}_{\left(R,n;C,m\right)} = \sum_ip_i {\rm vec}_{n}\left(\mathcal{T}^{(r_1\cdots r_p)}_{i}\right) {\rm vec}_{m}^{\rm T}\left(\mathcal{T}^{(c_1\cdots c_q)}_i\right).
	\end{equation*}
\end{lemma}

Using the dimension reduction techniques, we construct the extended correlation tensor for $N$-partite states $\rho\in\mathcal{H}_{d_1} \otimes \mathcal{H}_{d_2} \otimes \cdots \otimes \mathcal{H}_{d_N}$ ($N\geq2$) under a bipartition, which takes full account of all reduced density matrices and the correlation tensor. Given a bipartition $l_1\cdots l_k|l_{k+1}\cdots l_N$, where $\left\{l_1,\cdots, l_N\right\}=[N]$, $l_1<\cdots<l_k$, $l_{k+1}<\cdots<l_N$, $l_1<l_{k+1}$, $k\in[N-1]$, for any vectors $\bm{u}=\left(u_1,\cdots,u_{p_1}\right)^{\rm T}\in\mathbb{R}^{p_1}$, $\bm{v}=\left(v_1,\cdots,v_{p_2}\right)^{\rm T}\in\mathbb{R}^{p_2}$, $\bm{\alpha}=\left(a_1, \cdots, a_{q_1}\right)^{\rm T}\in\mathbb{R}^{q_1}$, and $\bm{\beta}=\left(b_1, \cdots, b_{q_2}\right)^{\rm T}\in\mathbb{R}^{q_2}$, the extended correlation tensor of $\rho$ under the bipartition has a size of $\left(p_1+q_1\left(d_{l_1}^2-1\right)\cdots\left(d_{l_k}^2-1\right)\right) \times \left(p_2+q_2\left(d_{l_{k+1}}^2-1\right)\cdots\left(d_{l_N}^2-1\right)\right)$ and is of the form, 
\begin{equation}\label{extended correlation tensor n-partite}
	\begin{aligned}
		& \mathcal{M}^{(l_1\cdots l_k|l_{k+1}\cdots l_{N})}_{\bm{u},\bm{v},\bm{\alpha},\bm{\beta}}\left(\rho\right) \\
		=& \begin{pmatrix}
			\bm{u}\bm{v}^{\rm T} & \bm{u}\left(\bm{\beta}\otimes{\rm vec}_{m}^{\rm T}\left(\mathcal{T}^{(l_{k+1}\cdots l_{N})}\right)\right) \\
			\left(\bm{\alpha}\otimes{\rm vec}_{n}\left(\mathcal{T}^{(l_{1}\cdots l_{k})}\right)\right)\bm{v}^{\rm T} & \bm{\alpha}\bm{\beta}^{\rm T}\otimes \mathbf{T}^{(l_{1}\cdots l_{k}l_{k+1}\cdots l_{N})}_{\left(R,n;C,m\right)}
		\end{pmatrix},
	\end{aligned}
\end{equation}
where $R=\left\{l_1,\cdots,l_k\right\}$, $n\in[k]$, $C=\left\{l_{k+1},\cdots,l_{N}\right\}$, $m\in[N-k]$.

We present the following theorem of separability criterion under a given bipartition for $n$-partite quantum states, see proof in Appendix~\ref{section: proof of theorem 4}.

\begin{theorem}\label{lemma: multipartite separability criterion}
	Given a bipartition $l_1\cdots l_k|l_{k+1}\cdots l_N$, where $\left\{l_1,\cdots, l_N\right\}=[N]$, $l_1<\cdots<l_k$, $l_{k+1}<\cdots<l_N$, $l_1<l_{k+1}$, $k\in[N-1]$, if an $N$-partite quantum state $\rho\in \mathcal{H}_{d_1} \otimes \mathcal{H}_{d_2} \otimes \cdots \otimes \mathcal{H}_{d_N}$ is biseparable under the bipartition $l_1\cdots l_k|l_{k+1}\cdots l_N$, then its extended correlation tensor \eqref{extended correlation tensor n-partite} satisfies that
	\begin{equation*}
		\left\|\mathcal{M}^{(l_1\cdots l_k|l_{k+1}\cdots l_N)}_{\bm{u},\bm{v},\bm{\alpha},\bm{\beta}}\left(\rho\right) \right\|_{tr} \leq \mathfrak{m}^{(l_1\cdots l_k|l_{k+1}\cdots l_N)}_{\bm{u},\bm{v},\bm{\alpha},\bm{\beta}},
	\end{equation*}
	where 
	\begin{equation*}
		\begin{aligned}
			& \mathfrak{m}^{(l_1\cdots l_k|l_{k+1}\cdots l_N)}_{\bm{u},\bm{v},\bm{\alpha},\bm{\beta}} \\
			=& \sqrt{\left\|\bm{u}\right\|_F^2+\left\|\bm{\alpha}\right\|_F^2\mathfrak{m}^{(l_1\cdots l_k)}} \sqrt{\left\|\bm{v}\right\|_F^2+\left\|\bm{\beta}\right\|_F^2\mathfrak{m}^{(l_{k+1}\cdots l_{N})}},
		\end{aligned}
	\end{equation*}
	$\mathfrak{m}^{(l_1\cdots l_k)}$ and $\mathfrak{m}^{(l_{k+1}\cdots l_N)}$ are defined in Eq.~\eqref{equation: m upper bound n-partite}.
\end{theorem}

For an $N$-partite quantum state $\rho\in \mathcal{H}_{d_1} \otimes\mathcal{H}_{d_2}\cdots\otimes\mathcal{H}_{d_N}$, denote
\begin{equation}\label{equation: M n-partite}
	\begin{aligned}
		& M^{(1\cdots N)}(\rho) \\
		=& \frac{1}{2^{N-1}-1} \sum_{k=1}^{N-1}\sum_{\substack{\left\{l_1,\cdots,l_N\right\}=[N]\\l_1<\cdots<l_k\\l_{k+1}<\cdots<l_N\\l_1<l_{k+1}}}\left\|\mathcal{M}_{\bm{u},\bm{v},\bm{\alpha},\bm{\beta}}^{(l_1\cdots l_k|l_{k+1}\cdots l_N)}(\rho)\right\|_{\rm tr},
	\end{aligned} 
\end{equation}
\begin{equation}\label{equation: M0 n-partite}
	M_0^{(1\cdots N)} = \sum_{k=1}^{N-1}\max_{\substack{\left\{l_1,\cdots,l_N\right\}=[N]\\l_1<\cdots<l_k\\l_{k+1}<\cdots<l_N\\l_1<l_{k+1}}} \left\{ \mathfrak{m}^{(l_1\cdots l_k|l_{k+1}\cdots l_N)}_{\bm{u},\bm{v},\bm{\alpha},\bm{\beta}} \right\}.
\end{equation}
Naturally, Theorem~\ref{lemma: multipartite separability criterion} produces the following theorem for the genuine $N$-partite entanglement, see proof in Appendix~\ref{section: proof of theorem 5}.

\begin{theorem}\label{theorem: genuine n-partite entangled}
	A quantum state $\rho\in\mathcal{H}_{d_1} \otimes \mathcal{H}_{d_2} \otimes \cdots \otimes \mathcal{H}_{d_N}$ is genuinely $N$-partite entangled if $M^{(1\cdots N)} > M^{(1\cdots N)}_0$.
\end{theorem}

We provide the following example of a tripartite quantum state to illustrate the Theorem~\ref{theorem: genuine n-partite entangled}.
\begin{example}
	Consider the $3\times3\times2$ Werner-type state  
	\begin{equation}\label{equation: state example 3}
		\rho_p = \frac{1-p}{18}\mathbf{I}_{18} + p \ket{\varphi}\bra{\varphi},
	\end{equation}
	where $\ket{\varphi} = \frac{1}{\sqrt{5}} \left( \left(\ket{10}+\ket{21}\right)\ket{0} + \left(\ket{00}+\ket{11}+\ket{22}\right)\ket{1} \right)$ is an entangled state~\cite{cornelio2006classification}, $0\leq p \leq1$.
	
	Choosing the Heisenberg-Weyl operators as the bases, we have $d_1=\kappa_1=3$, $d_2=\kappa_2=3$ and $d_3=\kappa_3=2$. Taking $\bm{u}=\left(x,x\right)^{\rm T}$, $\bm{v}=\left(\sqrt{x},\sqrt{x}\right)^{\rm T}$, $\bm{\alpha}=\left(-0.516382, -0.165015\right)^{\rm T}$ and $\bm{\beta}=\left(-0.259148, 0.485242\right)^{\rm T}$, by Theorem~\ref{theorem: genuine n-partite entangled} we have that $\rho_p$ is genuinely tripartite entangled for $0.233233\leq p\leq 1$. The result is shown in Figure~\ref{fig: example 5}, where the blue plane surface represents zero plane, the yellow curved surface represents the value of $M^{(123)}\left(\rho_{p}\right)-M^{(123)}_0$. 
	\begin{figure}[h!]
		\centering
		\includegraphics[width=0.5\textwidth]{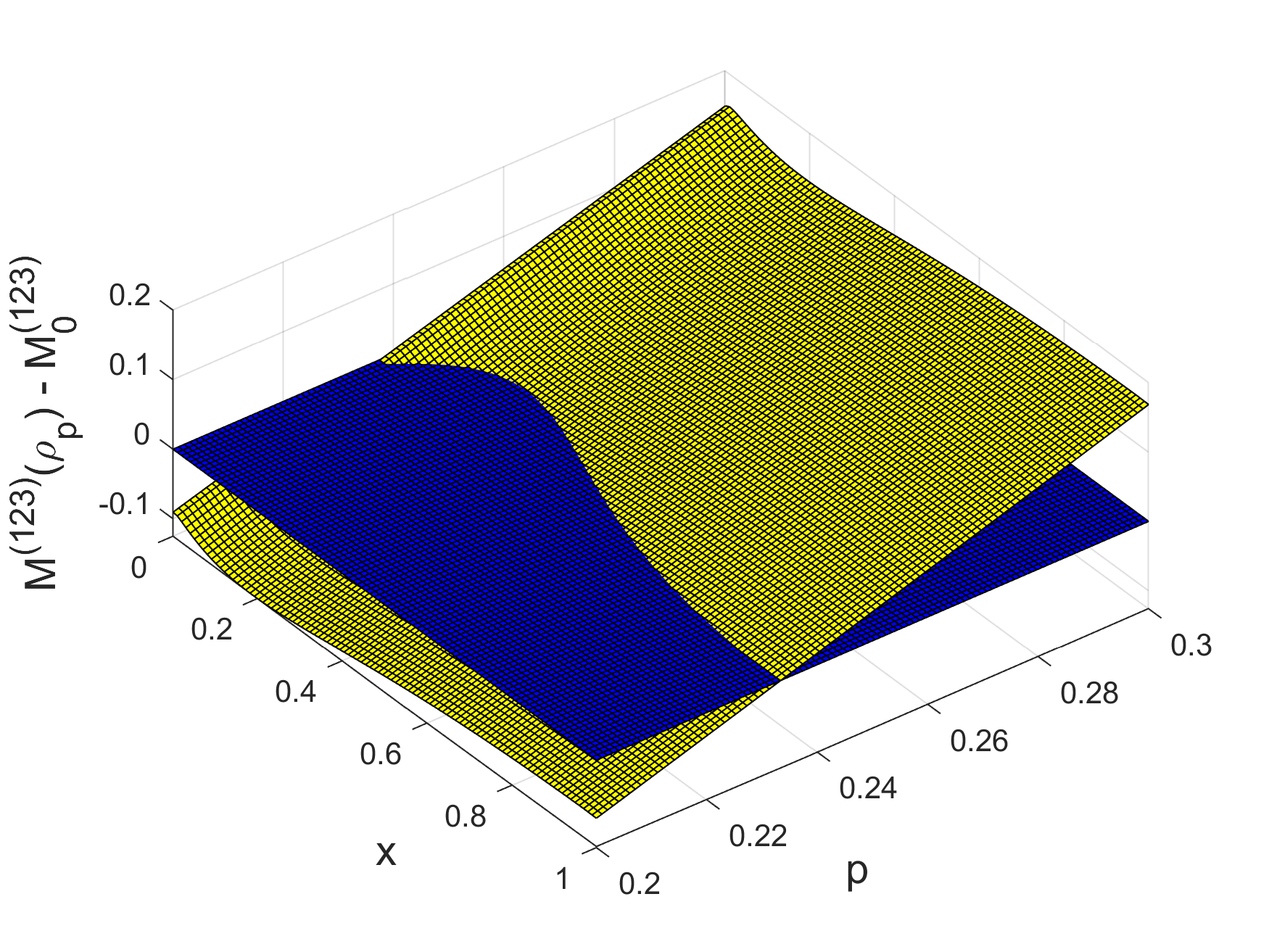}
		\caption{The relation between the entanglement detection of $\rho_p$ and the parameters $p$ and $x$ in Example 5 with $\bm{u}=\left(x,x\right)^{\rm T}$, $\bm{v}=\left(\sqrt{x},\sqrt{x}\right)^{\rm T}$, $\bm{\alpha}=\left(-0.516382, -0.165015\right)^{\rm T}$ and $\bm{\beta}=\left(-0.259148, 0.485242\right)^{\rm T}$.}
		\label{fig: example 5}
	\end{figure}
	
	We compare our result with those of Ref.~\cite{zhao2022detection,zhang2023improved} in Table~\ref{table: example 5}, which shows that our criterion detects a wider genuine-tripartite-entanglement region.
	\begin{table}[h!]
		\centering
		\caption{The comparison of detection thresholds for various criteria for the state $\rho_p$ in Eq.~\eqref{equation: state example 3}.}
		\begin{tabular}{>{\centering\arraybackslash}m{3cm}>{\centering\arraybackslash}m{4cm}}
			\hline
			\hline
			\textbf{Criteria} & \textbf{Range of $p$} \\
			\hline
			Theorem~\ref{theorem: genuine n-partite entangled} & $0.233233\leq p\leq1$ \\
			Zhang's criterion~\cite{zhang2023improved} & $0.51< p\leq1$ \\
			Zhao's criterion~\cite{zhao2022detection} & $0.5635<p\leq1$ \\
			\hline
			\hline
		\end{tabular}
		\label{table: example 5}
	\end{table}   
\end{example}

Concerning  the fully separablity of $N$-partite states, we have the following lemma, see proof in Appendix~\ref{section: proof of lemma 4}.

\begin{lemma}\label{lemma: property of correlation tensor n fully separable}
	If an $N$-partite quantum state $\rho\in\mathcal{H}_{d_1}\otimes\cdots\otimes\mathcal{H}_{d_N}$ is fully separable, 
	$\rho = \sum_ip_i\rho^{(1)}_i \otimes \cdots\otimes\rho^{(N)}_i$, where $0\leq p_i\leq 1$, $\sum_ip_i=1$, $\rho^{(k)}_i$ is the pure state in the subsystem $\mathcal{H}_{d_k}$, then for any $\left\{r_1,\cdots,r_k\right\}\subseteq[N]$, the correlation tensor $\mathcal{T}^{(r_1\cdots r_k)}$ of the reduced state $\rho^{(r_1\cdots r_k)}$ can be expressed as
	$\mathcal{T}^{(r_1\cdots r_k)} = \sum_ip_i \mathcal{T}^{(r_1)}_i\circ\cdots\circ\mathcal{T}^{(r_k)}_i$, where $\mathcal{T}^{(r_l)}_i$ is the correlation tensor of the reduced state $\rho^{(r_l)}_i$, $l\in[k]$. 
\end{lemma}

The following corollary provides a criterion to detect the fully separability for multipartite states, which generalizes the Theorem 1 in Ref.~\cite{hassan2008separability}, see proof in Appendix~\ref{section: proof of theorem 6}.
\begin{corollary}\label{theorem: fully separability}
	If an $N$-partite quantum state $\rho\in\mathcal{H}_{d_1}\otimes\mathcal{H}_{d_2}\otimes\cdots\otimes\mathcal{H}_{d_N}$ is fully separable, then
	\begin{equation*}
		\left\|\mathcal{T}^{(1\cdots N)}\right\|_{\rm tr} \leq \prod_{i=1}^{N} \sqrt{\frac{d_i^2-d_i}{\kappa_i}}.
	\end{equation*}
\end{corollary}

\begin{example}\label{example: ghz perturbation}
	Consider the following state \cite{gittsovich2010multipartite},
	\begin{equation}\label{equation: state ghz perturbation}
		\rho_{p,\epsilon} = \frac{1-p}{8} + p\ket{\psi_{\epsilon}}\bra{\psi_\epsilon},
	\end{equation}
	where $0\leq p \leq 1$,  $\ket{\psi_\epsilon} = \frac{1}{\sqrt{2+\epsilon^2}} \left(\ket{000} + \epsilon\ket{110} + \ket{111}\right)$ with real parameter $\epsilon$.
	The dimensions of this state are $d_1=d_2=d_3=2$. Choosing the Pauli operators as the bases gives that $\kappa_1=\kappa_2=\kappa_3=2$. Corollary~\ref{theorem: fully separability} shows that the state $\rho_{p,\epsilon}$ is entangled for the ranges of $p$ and $\epsilon$ shown in Figure~\ref{fig: example 6}, where the blue plane surface represents the zero plane, and the yellow curved surface represents the difference of the trace norm and the upper bound given in Corollary~\ref{theorem: fully separability}. 
	\begin{figure}[h!]
		\centering
		\includegraphics[width=0.5\textwidth]{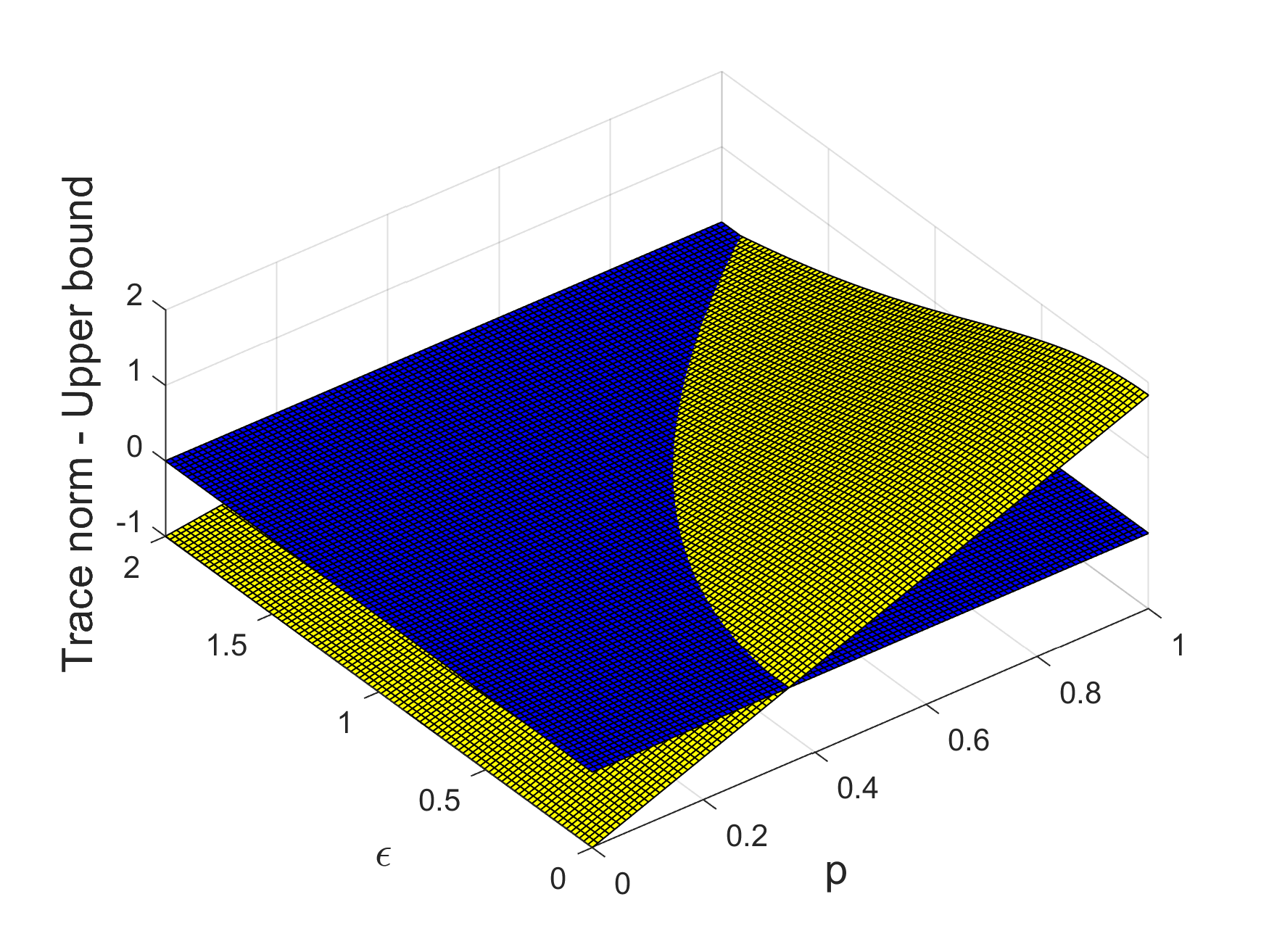}
		\caption{The relation between the entanglement detection of $\rho_{p,\epsilon}$ and the parameters $p$ and $\epsilon$ in Example~\ref{example: ghz perturbation}.}
		\label{fig: example 6}
	\end{figure}
	
	We compare our criterion with those in Ref.~\cite{zhang2017realignment,shen2022optimization,sun2025separability} for $\epsilon = 10^{-1}$ and $\epsilon = 1$ in Table~\ref{table: example 6}, which implies that our criterion detects a wider range of entanglement for small $\epsilon$.
	\begin{table}[ht!]
		\centering
		\caption{The comparison of detection thresholds for various criteria for the state $\rho_{p,\epsilon}$ in Eq.~\eqref{equation: state ghz perturbation}.}
		\begin{tabular}{>{\centering\arraybackslash}m{2.7cm}>{\centering\arraybackslash}m{2.3cm}>{\centering\arraybackslash}m{2.2cm}}
			\hline\hline
			$\epsilon$ & $10^{-1}$ & $1$ \\
			\hline
			Zhang's criterion~\cite{zhang2017realignment} & $0.4118\leq p\leq1$ & $0.4256< p\leq1$ \\
			Shen's criterion~\cite{shen2022optimization} & $0.4039\leq p\leq1$ & $0.4200<p\leq1$ \\
			Sun's criterion~\cite{sun2025separability} & $0.4026\leq p \leq 1$ & $0.4194\leq p \leq 1$ \\
			Corollary~\ref{theorem: fully separability} & $0.355322\leq p\leq1$ & $0.53034\leq p\leq 1$ \\
			\hline\hline
		\end{tabular}
		\label{table: example 6}
	\end{table}
\end{example}

\section{Conclusions}\label{section: conclusion}
In this work, we have established a unified and versatile framework for detecting entanglements in bipartite and multipartite quantum systems. At the core of this framework lies a novel extended correlation tensor constructed for bipartite states using the generalized Bloch representation, which incorporates adjustable parameters and allows flexibility in the choice of orthogonal basis. This design not only enhances the generality of the separability criterion but also bridges several existing criteria, such as those by de Vicente~\cite{devicente2007separability}, Shen et al.~\cite{shen2016improved}, Sarbicki et al.~\cite{sarbicki2020family}, Zhu et al.~\cite{zhu2023family}, and Huang et al.~\cite{huang2024unifying}, as special cases of our unified result.

To generalize the framework to multipartite systems, we adopt the technique of generalized matricization, specifically $\left(R,n;C,m\right)$-mode matrix unfolding, which enables the systematic construction of extended correlation tensors under arbitrary bipartitions. This enabled us to derive practical criteria for both biseparability and full separability in multipartite states. Numerical experiments validated the effectiveness of our criteria, demonstrating enhanced detection capability over existing methods. Future works might be focused on establishing $k$-separability criteria for general multipartite systems.

\section*{Declaration of competing interest}
The authors declare no conflict of interest.
	
\section*{Data availability}
No data was used for the research described in the article.

\section*{Acknowledgments}
This work is supported by the Stable Supporting Fund of Acoustic Science and Technology Laboratory (Grant No. JCKYS2024604SSJS001) and the Fundamental Research Funds for the Central Universities (Grant No. 3072024XX2401).

\appendix

\section{Several orthogonal basis}\label{section: several orthogonal basis}
In this section, several orthogonal bases are presented, including Pauli operators, Weyl operators, and Heisenberg-Weyl operators.

Four $2$-dimensional Pauli matrices are denoted as $\bm{\sigma}_0 = \begin{pmatrix}
    1 & 0 \\ 
    0 & 1
\end{pmatrix}$,
$\bm{\sigma}_1=\begin{pmatrix}
    0 & 1 \\
    1 & 0
\end{pmatrix}$,
$\bm{\sigma}_2=\begin{pmatrix}
    0 & -i \\
    i & 0
\end{pmatrix}$,
$\bm{\sigma}_3=\begin{pmatrix}
    1 & 0 \\
    0 & -1
\end{pmatrix}$.
And the explicit matrix representations of Pauli operators on the $2^N$-dimensional Hilbert space are the tensor product of $N$ $2$-dimensional Pauli matrices.

Let $\mathcal{H}$ be a $d$-dimensional Hilbert space with computational basis $\left\{\ket{k}\right\}$, and $\mathbb{Z}_d$ denote the finite field of modulo $d$ integers. The Weyl operators are defined by $\mathbf{W}\left(n,m\right) = \sum_{k\in\mathbb{Z}_d} e^{\frac{2kn\pi i}{d}}\ket{k}\bra{(k+m)\text{mod }d}$,
where $n,m\in[0,d-1]$. The set $\left\{\mathbf{W}\left(n,m\right)\right\}$ forms the basis for linear generators in the general linear Lie algebra $\mathfrak{gl}(d)$~\cite{huang2022separability}. If $d=2$, the Weyl operators are equal to the $2$-dimensional Pauli operators. 

The $d$-dimensional Heisenberg-Weyl operators~\cite{asadian2016heisenberg,chang2018separability} are defined by $\mathbf{Q}\left(l,m\right) = \chi \mathbf{W}\left(n,m\right) + \chi^*\mathbf{W}^{\dagger}\left(l,m\right)$,
where $\chi=\left(1+i\right)/2$.

\section{Matrix unfolding of tensors}
In special cases, the ${(R,n;C,m)}$-mode matrix unfolding can reduce to the vectorization or the matrix unfolding of Ref.~\cite{delathauwer2000multilinear}, as shown in Table~\ref{tab: relation between matrization and vectorization}.
\begin{table}[htbp]
    \centering
    \begin{tabular}{cccc|c}
        \hline\hline 
        $R$ & $n$ & $C$ & $m$ & $\mathbf{A}_{(R,n;C,m)}$ \\
        \hline
         $[N]$ & $n$ & $\emptyset$ & / & $\operatorname{vec}_n\left(\mathcal{A}\right)$ \\ 
         $\emptyset$ & / & $[N]$ & $m$ & $\operatorname{vec}_m^{\rm T}\left(\mathcal{A}\right)$ \\
         $\left\{r_1\right\}$ & $1$ & $[N]\setminus \left\{r_1\right\}$ & $N-1$ & $\mathbf{A}_{(r_1)}$ \\
         $[N]\setminus \left\{c_1\right\}$ & $N-1$ & $\left\{c_1\right\}$ & $1$ & $\mathbf{A}_{(c_1)}^{\rm T}$ \\
        \hline\hline 
    \end{tabular}
    \caption{The reduction from the $(R,n;C,m)$-mode matrix unfolding to the vectorization or the matrix unfolding of Ref.~\cite{delathauwer2000multilinear} for a tensor $\mathcal{A}\in\mathbb{C}^{d_1\times\cdots\times d_N}$ in special cases.}
    \label{tab: relation between matrization and vectorization}
\end{table}

The process of the $\left(R,n;C,m\right)$-mode matrix unfolding of tensors can be visually decomposed into the following three steps:
\begin{itemize}
	\item \textbf{Step 1:} A tensor $\mathcal{A}=\left(a_{i_1\cdots i_N}\right)\in\mathbb{C}^{d_1\times\cdots\times d_N}$ can be considered as a special ``tensor'' $\tilde{\mathcal{A}}= \left(\tilde{a}_{i_{r_1}\cdots i_{r_k}}\right)$ of size $d_{r_1}\times\cdots\times d_{r_k}$, whose entries $\tilde{a}_{i_{r_1}\cdots i_{r_k}}$ are sub-tensors of size $d_{c_1}\times\cdots\times d_{c_{N-k}}$.
	
	\item \textbf{Step 2:} Performing the $n$-mode vectorization to $\tilde{\mathcal{A}}$ yields a ``column vector'' ${\rm vec}_{n}\left(\tilde{\mathcal{A}}\right)$, whose entries $\tilde{a}_i$ have size $d_{c_1}\times\cdots\times d_{c_{N-k}}$ for $i\in[d_{r_1}\cdots d_{r_k}]$.
	
	\item \textbf{Step 3:} Perform the $m$-mode vectorization to all entries $\tilde{a}_j$ and then transpose them, respectively.
\end{itemize}

During the expansion process, the rows are first processed, followed by the columns. Alternatively, $\mathcal{A}$ can be viewed as a special ``tensor'' $\tilde{\mathcal{A}}'$ of size $d_{c_1}\times\cdots\times d_{c_{N-k}}$, where each entry is a sub-tensor of size $d_{r_1}\times\cdots\times d_{r_{k}}$. In this alternative approach, the columns are processed first, followed by the rows. Both approaches yield the same matrix.

The following theorem gives the relation between the $\left(R,n;C,m\right)$-mode matrix unfolding of tensors with the outer product.
\begin{theorem}\label{theorem: relation between matrix unfolding and outer product}
	Given two tensors $\mathcal{A}\in\mathbb{C}^{d_{1}\times\cdots\times d_{k}}$, $\mathcal{B}\in\mathbb{C}^{d_{k+1}\times\cdots\times d_{N}}$ with $k\in[N-1]$, let $\mathcal{C} = \mathcal{A}\circ\mathcal{B}$ and $R=\left\{r_1,\cdots,r_k\right\}, C=\left\{c_1,\cdots,c_{N-k}\right\}$ be two disjoint ascending ordered sets with $R\cup C=[N]$. Then, for any $n\in[k]$ and $m\in[N-k]$, we have 
    $$\mathbf{C}_{\left(R,n;C,m\right)} = {\rm vec}_{n}\left(\mathcal{A}\right)  {\rm vec}_{m}^{\rm T}\left(\mathcal{B}\right).$$
\end{theorem}
\begin{proof}
    Let $\mathcal{A}=\sum_{i_1,\cdots,i_k} a_{i_1\cdots i_k} \bm{e}^{(1)}_{i_1}\circ\cdots\circ\bm{e}^{(k)}_{i_k}$ and $\mathcal{B}=\sum_{i_{k+1},\cdots,i_N} b_{i_{k+1}\cdots i_N} \bm{e}^{(k+1)}_{i_{k+1}}\circ\cdots\circ\bm{e}^{(N)}_{i_N}$. Then $\mathcal{C}=\mathcal{A}\circ\mathcal{B}=\sum_{i_1,\cdots,i_N} a_{i_1\cdots i_k}b_{i_{k+1}\cdots i_N} \bm{e}^{(1)}_{i_1}\circ\cdots\circ\bm{e}^{(N)}_{i_N}$. Therefore, we have
    \begin{equation*}
        \begin{aligned}
            &\mathbf{C}_{\left(R,n;C,m\right)} \\
            =& \sum_{i_1,\cdots,i_N} a_{i_1\cdots i_k}b_{i_{k+1}\cdots i_N} \left(\left(\bigotimes_{i=n+1}^k \bm{e}^{(i)}\right)\otimes\left(\bigotimes_{i=1}^n \bm{e}^{(i)}\right)\right) \\
            & \cdot \left(\left(\bigotimes_{i=m+1}^{N-k} \bm{e}^{(k+i)}\right)\otimes\left(\bigotimes_{i=1}^{m} \bm{e}^{(k+i)}\right)\right)^{\rm T} \\
            =& \left(\sum_{i_1,\cdots,i_k} a_{i_1\cdots i_k} \left(\bigotimes_{i=n+1}^k \bm{e}^{(i)}\right)\otimes\left(\bigotimes_{i=1}^n \bm{e}^{(i)}\right)\right) \\
            & \cdot \left(\sum_{i_{k+1},\cdots,i_N} b_{i_{k+1}\cdots i_N} \left(\bigotimes_{i=m+1}^{N-k} \bm{e}^{(k+i)}\right)\otimes\left(\bigotimes_{i=1}^{m} \bm{e}^{(k+i)}\right)\right)^{\rm T} \\
            =& \operatorname{vec}_n\left(\mathcal{A}\right) \operatorname{vec}_m^{\rm T}\left(\mathcal{B}\right)
        \end{aligned}
    \end{equation*}
\end{proof}

\section{Proof of Lemmas and Theorems}\label{section: proof of lemmas and theorems}
\subsection{Proof of Theorem~\ref{theorem: bipartite separability criteria}}\label{section: proof of theorem 1}
\begin{proof}
	Since $\rho^{(AB)}$ is separable, it can be expressed as a convex combination of product states, 
	\begin{equation}\label{equation: convex combination bipartite}
		\rho^{(AB)} = \sum_i p_i \rho^{(A)}_{i} \otimes \rho^{(B)}_{i},
	\end{equation} 
	where $0\leq p_i\leq1$, $\sum_{i}p_i=1$, $\rho^{(A)}_{i}$ and $\rho^{(B)}_{i}$ are pure states in $\mathcal{H}_{d_A}$ and $\mathcal{H}_{d_B}$, respectively. 
	
	By the generalized Bloch representation of $\rho^{(A)}_{i},\rho^{(B)}_{i}$ in Eq~\eqref{generalized bloch representation}, there exist vectors $\mathcal{T}^{(A)}_i\in\mathbb{C}^{d_A^2-1}$, $\mathcal{T}^{(B)}_i\in\mathbb{C}^{d_B^2-1}$ 
	such that $\mathcal{T}^{(A)}=\sum_i p_i \mathcal{T}^{(A)}_i$, $\mathcal{T}^{(B)}=\sum_i p_i \mathcal{T}^{(B)}_i$. According to Lemma~\ref{lemma: bloch vector upper bound}, we have $\left\|\mathcal{T}^{(A)}_i\right\|_F^2\leq\frac{d_A^2-d_A}{\kappa_A}$, $\left\|\mathcal{T}^{(B)}_i\right\|_F^2\leq\frac{d_B^2-d_B}{\kappa_B}$.
	
	Substituting the generalized Bloch representation of $\rho^{(AB)}$ in Eq.~\eqref{generalized bloch representation bipartite} and the generalized Bloch representations of $\rho^{(A)}_{i},\rho^{(B)}_{i}$ in Eq.~\eqref{generalized bloch representation} into Eq.~\eqref{equation: convex combination bipartite}, we derive that $\mathcal{T}^{(AB)} = \sum_i{p_i}\mathcal{T}^{(A)}_i\left(\mathcal{T}^{(B)}_i\right)^{\rm T}$.
	
	Therefore, we obtain that
	\begin{equation*}
		\begin{aligned}
			&\left\| \mathcal{M}_{\bm{u},\bm{v},\bm{\alpha},\bm{\beta}}^{(A|B)}(\rho^{(AB)}) \right\|_{\rm tr} \\
			=&\left\|\begin{pmatrix}
				\bm{u}\bm{v}^{\rm T} & \bm{u}\left(\bm{\beta}\otimes \mathcal{T}^{(B)}\right)^{\rm T} \\
				\left(\bm{\alpha} \otimes \mathcal{T}^{(A)}\right)\bm{v}^{\rm T} & \bm{\alpha}\bm{\beta}^{\rm T} \otimes \mathcal{T}^{(AB)}\\
			\end{pmatrix}\right\|_{\rm tr} \\
			=&\left\|\begin{pmatrix}
				\bm{u}\bm{v}^{\rm T} & \bm{u}\left(\bm{\beta}\otimes \sum_i p_i \mathcal{T}^{(B)}_i\right)^{\rm T} \\
				\left(\bm{\alpha} \otimes \sum_i p_i \mathcal{T}^{(A)}_i\right)\bm{v}^{\rm T} & \bm{\alpha}\bm{\beta}^{\rm T} \otimes \left(\sum_i{p_i}\mathcal{T}^{(A)}_i\left(\mathcal{T}^{(B)}_i\right)^{\rm T}\right) \\
			\end{pmatrix}\right\|_{\rm tr} \\
			\leq& \sum_ip_i\left\|\begin{pmatrix}
				\bm{u}\bm{v}^{\rm T} & \bm{u}\left(\bm{\beta}\otimes \mathcal{T}^{(B)}_i\right)^{\rm T} \\
				\left(\bm{\alpha} \otimes \mathcal{T}^{(A)}_i\right)\bm{v}^{\rm T} & \bm{\alpha}\bm{\beta}^{\rm T} \otimes \left(\mathcal{T}^{(A)}_i\left(\mathcal{T}^{(B)}_i\right)^{\rm T}\right) \\
			\end{pmatrix}\right\|_{\rm tr} \\
			\leq&\sum{p_i}\left\|\begin{pmatrix}
				\bm{u} \\
				\bm{\alpha} \otimes \mathcal{T}^{(A)}_i \\
			\end{pmatrix}\right\|_{\rm tr}
			\left\|\begin{pmatrix}
				\bm{v}^{\rm T} & \left(\bm{\beta}\otimes \mathcal{T}^{(B)}_i\right)^{\rm T}
			\end{pmatrix}\right\|_{\rm tr} \\
			\leq& \sqrt{\left(\left\|\bm{u}\right\|_F^2+\left\|\bm{\alpha}\right\|_F^2\frac{{d_A}^2-d_A}{ \kappa_A}\right) \left(\left\|\bm{v}\right\|_F^2+\left\|\bm{\beta}\right\|_F^2\frac{{d_B}^2-d_B}{ \kappa_B}\right)}.
		\end{aligned}
	\end{equation*}      
\end{proof}

\subsection{Proof of Lemma~\ref{lemma: upper bound of correlation tensor n-paritite}}\label{section: proof of lemma 2}
\begin{proof}
Denote $\rho^{(1\cdots N)}$ the state in $\mathcal{H}_{d_1}\otimes\cdots\otimes\mathcal{H}_{d_N}$ of the multipartite system. By the generalized Bloch representation in Eq.~\eqref{generalized bloch representation n} and the condition in Eq.~\eqref{equation: base condition n-partite}, we obtain that
\begin{equation}\label{equation: tr(rho)}
	\begin{aligned}
		&{\rm Tr}\left(\left(\rho^{(1\cdots N)}\right)^2\right) \\
		=& {\rm Tr}\left(\left(\rho^{(1\cdots N)}\right)^{\dagger}\rho^{(1\cdots N)}\right)\\
		=& \frac{1}{d_1^2\cdots d_N^2}\left(d_1\cdots d_N \right. \\
		& +\kappa_1d_2\cdots d_N\left\|\mathcal{T}^{(1)}\right\|_F^2+\cdots +\kappa_Nd_1\cdots d_{N-1}\left\|\mathcal{T}^{(N)}\right\|_F^2  \\
		& +\kappa_1\kappa_2d_3\cdots d_N\left\|\mathcal{T}^{(12)}\right\|_F^2
		+\cdots\\
		& +\kappa_{N-1}\kappa_Nd_1\cdots d_{N-2}\left\|\mathcal{T}^{(N-1N)}\right\|_F^2 \\
		& \left.+\cdots +\kappa_1\cdots\kappa_N\left\|\mathcal{T}^{(1\cdots N)}\right\|_F^2\right) \\
		=& \frac{1}{d_1\cdots d_N}\left(1+A_1^{(1\cdots N)}+A_2^{(1\cdots N)}+\cdots+A_N^{(1\cdots N)}\right),
	\end{aligned}
\end{equation}
where
\begin{equation*}
	\begin{gathered}
		A_1^{(1\cdots N)}=\frac{\kappa_1}{d_1}\left\|\mathcal{T}^{(1)}\right\|_F^2+\cdots+\frac{\kappa_N}{d_N}\left\|\mathcal{T}^{(N)}\right\|_F^2, \\
		A_2^{(1\cdots N)}=\frac{\kappa_1\kappa_2}{d_1d_2}\left\|\mathcal{T}^{(12)}\right\|_F^2+\cdots+\frac{\kappa_{N-1}\kappa_N}{d_{N-1}d_N}\left\|\mathcal{T}^{(N-1N)}\right\|_F^2, \\
		\cdots,\\
		A_n^{(1\cdots N)}=\frac{\kappa_1\cdots\kappa_N}{d_1\cdots d_N}\left\|\mathcal{T}^{(1\cdots N)}\right\|_F^2.
	\end{gathered}
\end{equation*}

Let $\rho^{(j_1)}$ and $\rho^{(j_2\cdots j_N)}$ be the reduced states of $\rho^{(1\cdots N)}$ with respect to the subsystems $\mathcal{H}_{d_{j_1}}$ and $\mathcal{H}_{d_{j_2}}\otimes\cdots\otimes\mathcal{H}_{d_{j_N}}$, respectively, where $\left\{j_1, j_2,\cdots, j_N\right\}=[N]$. 

If $\rho^{(1\cdots N)}$ is pure state, we have 
\begin{equation}\label{equation: Tr(rho^2)}
	\begin{aligned}
		& {\rm Tr}\left(\left(\rho^{(1\cdots N)}\right)^2\right) \\
		=& \frac{1}{d_1\cdots  d_N}\left(1+A_1^{(1\cdots N)}+A_2^{(1\cdots N)}+\cdots+A_N^{(1\cdots N)}\right)\\
		=& 1,
	\end{aligned}
\end{equation}
and 
\begin{equation}\label{equation: rhoj1 = rhoj2jN}
	{\rm Tr}\left(\left(\rho^{(j_1)}\right)^2\right)={\rm Tr}\left(\left(\rho^{(j_2\cdots j_N)}\right)^2\right),
\end{equation}
which follows that
\begin{equation}\label{equation: left and right}
	\sum_{j_1=1}^N\frac{1}{d_{j_1}}{\rm Tr}\left(\left(\rho^{(j_1)}\right)^2\right)=\sum_{j_1=1}^N\frac{1}{d_{j_1}}{\rm Tr}\left(\left(\rho^{(j_2\cdots j_N)}\right)^2\right).
\end{equation}

On one hand, $\rho^{(j_1)}$ is computed by performing the partial trace to $\rho^{(1\cdots N)}$, that is,
\begin{equation*}
	\rho^{(j_1)} = {\rm Tr}_{j_2\cdots j_N}\left(\rho^{(1\cdots N)}\right)=\frac{1}{d_{j_1}}\left(\mathbf{I}_{d_{j_1}}^{(j_1)}+\sum_{i_{j_1}=1}^{d_{j_1}^2-1}t_{i_{j_1}}^{(j_1)}\mathbf{G}_{i_{j_1}}^{(j_1)}\right).
\end{equation*}
Then we have
\begin{equation}\label{equation: left part}
	\begin{aligned}
		&\sum_{j_1=1}^N\frac{1}{d_{j_1}}{\rm Tr}\left(\left(\rho^{(j_1)}\right)^2\right) \\
		=& \sum_{j_1=1}^N\frac{1}{d_{j_1}}\left(\frac{1}{d_{j_1}^2} \left(d_{j_1}+\kappa_{j_1}\left\|\mathcal{T}^{(j_1)}\right\|_F^2\right)\right) \\
		=& \sum_{j_1=1}^N\frac{1}{d_{j_1}^2} + \sum_{j_1=1}^N\frac{\kappa_{j_1}}{d_{j_1}^3}\left\|\mathcal{T}^{(j_1)}\right\|_F^2.
	\end{aligned}
\end{equation}
On the other hand, according to Eq.~\eqref{equation: tr(rho)}, we obtain that
\begin{equation}\label{equation: right part}
	\begin{aligned}
		& \sum_{j_1=1}^N\frac{1}{d_{j_1}}{\rm Tr}\left(\left(\rho^{(j_2\cdots j_N)}\right)^2\right) \\
		=& \frac{1}{d_1}{\rm Tr}\left(\left(\rho^{(2\cdots N)}\right)^2\right) +\frac{1}{d_2}{\rm Tr}\left(\left(\rho^{(13\cdots N)}\right)^2\right) \\
		& +\cdots +\frac{1}{d_N}{\rm Tr}\left(\left(\rho^{(1\cdots N-1)}\right)^2\right) \\
		=& \frac{1}{d_1\cdots d_N} \left( N+ \left(A_1^{(2\cdots N)} + A_1^{(13\cdots N)} + \cdots + A_1^{(1\cdots N-1)}\right) \right. \\
		& + \left(A_2^{(2\cdots N)} + A_2^{(13\cdots N)} + \cdots + A_2^{(1\cdots N-1)}\right) \\
		& \left.+ \cdots + \left(A_{N-1}^{(2\cdots N)} + A_{N-1}^{(13\cdots N)} + \cdots + A_{N-1}^{(1\cdots N-1)}\right)\right) \\
		=& \frac{1}{d_1\cdots d_N}\left( N+ \left(N-1\right)A_1^{(1\cdots N)}+ \left(N-2\right)A_2^{(1\cdots N)}\right.\\
		& \left.+\cdots+A_{N-1}^{(1\cdots N)}\right).
	\end{aligned}
\end{equation}
Substituting Eq.~\eqref{equation: left part} and~\eqref{equation: right part} into Eq.~\eqref{equation: Tr(rho^2)}, we have
\begin{equation}
	\begin{aligned}
		& \sum_{i=1}^N\frac{1}{d_{i}^2} + \sum_{i=1}^N\frac{\kappa_{i}}{d_{i}^3}\left\|\mathcal{T}^{(i)}\right\|_F^2 \\
		=& \frac{1}{d_1\cdots d_N}\left( N+ \left(N-1\right)A_1^{(1\cdots N)}+ \left(N-2\right)A_2^{(1\cdots N)}\right.\\
		& \left.+\cdots+A_{N-1}^{(1\cdots N)}\right).
	\end{aligned}
\end{equation}

{\bf Case 1:} $N=1$. 

By Lemma~\ref{lemma: bloch vector upper bound}, we have $\left\|\mathcal{T}^{(1)}\right\|_F^2 \leq \frac{d_1^2-d_1}{\kappa_1}$.

{\bf Case 2:} $N=2$ or $N\geq3$, $\frac{d_1\cdots d_N}{d^2} < 1$. 

Since $\left\|\mathcal{T}^{(i)}\right\|_F^2\geq0$, we have
\begin{equation*}
	\begin{aligned}
		&\sum_{i=1}^N\frac{1}{d_{i}^2} \\
		\leq& \frac{1}{d_1\cdots d_N}\left( N+ \left(N-1\right)\left(A_1^{(1\cdots N)}+ A_2^{(1\cdots N)} \right.\right. \\
		& \left.\left.+\cdots+A_{N-1}^{(1\cdots N)}\right)\right) \\
		=& \frac{1}{d_1\cdots d_N}\left( N+ \left(N-1\right)\left(d_1\cdots d_N -1 - A_{N}^{(1\cdots N)}\right)\right),
	\end{aligned}
\end{equation*}
that is,
\begin{equation*}
	\begin{aligned}
		& \left\|\mathcal{T}^{(1\cdots N)}\right\|_F^2 \\
		\leq& \frac{d_1\cdots d_N}{\kappa_1\cdots \kappa_N} \left(d_1\cdots d_N + \frac{1}{N-1} - \frac{d_1\cdots d_N}{N-1}\sum_{i=1}^N\frac{1}{d_i^2}\right).
	\end{aligned}
\end{equation*}

{\bf Case 3:} $N\geq3$, $\frac{d_1\cdots d_N}{d^2} \geq 1$.

Since $d=\max_{i\in[N]}\left\{d_i\right\}$ and $\frac{d_1\cdots d_N}{d^2}\geq1$, we have
\begin{equation*}
	\begin{aligned}
		&\sum_{j_1=1}^N\frac{1}{d_{j_1}^2} + \sum_{j_1=1}^N\frac{\kappa_{j_1}}{d_{j_1}^3}\left\|\mathcal{T}^{(j_1)}\right\|_F^2 \\
		\geq& \sum_{j_1=1}^N\frac{1}{d_{j_1}^2} + \frac{1}{d^2}\sum_{j_1=1}^N\frac{\kappa_{j_1}}{d_{j_1}}\left\|\mathcal{T}^{(j_1)}\right\|_F^2 \\
		\geq& \sum_{j_1=1}^N\frac{1}{d_{j_1}^2} + \frac{1}{d_1\cdots d_N} A^{(1\cdots N)}_1,
	\end{aligned}
\end{equation*}
Then we obtain that
\begin{equation*}
	\begin{aligned}
		& \sum_{j_1=1}^N\frac{1}{d_{j_1}^2} + \frac{1}{d_1\cdots d_N} A^{(1\cdots N)}_1 \\
		\leq& \frac{1}{d_1\cdots d_N}\left( N+ \left(N-1\right)A_1^{(1\cdots N)} \right. \\
		& \left.+ \left(N-2\right)\left(A_2^{(1\cdots N)}+\cdots+A_{N-1}^{(1\cdots N)}\right)\right),
	\end{aligned}
\end{equation*}
that is,
\begin{equation*}
	\begin{aligned}
		& \frac{d_1\cdots d_N}{N-2}\sum_{j_1=1}^N\frac{1}{d_{j_1}^2} \\
		\leq& \frac{N}{N-2} + A_1^{(1\cdots N)} + A_2^{(1\cdots N)}+\cdots+A_{N-1}^{(1\cdots N)} \\
		=& \frac{2}{N-2} + d_1\cdots d_N - A^{(1\cdots N)}_N,
	\end{aligned}
\end{equation*}
which follows that
\begin{equation*}
	\begin{aligned}
		& \left\|\mathcal{T}^{(1\cdots N)}\right\|_F^2 \\
		\leq& \frac{d_1\cdots d_N}{\kappa_1\cdots \kappa_N}\left(d_1\cdots d_N+\frac{2}{N-2}-\frac{d_1\cdots d_N}{N-2}\sum_{i=1}^N
		\frac{1}{d_i^2}\right).
	\end{aligned}
\end{equation*}

Thus, by the above three cases, it holds that
\begin{equation*}
	\left\|\mathcal{T}^{(1\cdots N)}\right\|_F^2 \leq \mathfrak{m}^{(1\cdots N)}.
\end{equation*}

If $\rho^{(1\cdots N)}$ is a mixed state, then it can be represented as a convex combination of pure states, that is, $\rho^{(1\cdots N)}=\sum_i p_i\rho_{i}^{(1\cdots N)}$, where $0\leq p_i\leq1$, $\sum_ip_i=1$. Therefore, by the convexity of the Frobenius norm, we obtain that
\begin{equation*}
	\begin{aligned}
		\left\|\mathcal{T}^{(1\cdots N)}\right\|_F^2 
		= \left\|\sum_ip_i\mathcal{T}^{(1\cdots N)}_i\right\|_F^2
		\leq& \sum_ip_i\left\|\mathcal{T}^{(1\cdots N)}_i\right\|_F^2 \\
		\leq& \mathfrak{m}^{(1\cdots N)},
	\end{aligned}
\end{equation*}
where $\mathcal{T}^{(1\cdots N)}_i$ is the correlation tensor of $\rho_{i}^{(1\cdots N)}$.
\end{proof}

\subsection{Proof of Lemma~\ref{lemma: property of correlation tensor n}}\label{section: proof of lemma 3}
\begin{proof}
Suppose that $\left\{l_1,\cdots,l_k\right\}\setminus \left\{r_1,\cdots, r_p\right\} = \left\{r_{p+1},\cdots,r_k\right\}$ and $\left\{l_{k+1},\cdots,l_N\right\} \setminus \left\{c_1,\cdots, c_q\right\} = \left\{c_{q+1},\cdots,c_{N-k}\right\}$. Then we have
\begin{equation}\label{equation: rho_rs}
	\begin{aligned}
		& \rho^{(r_1\cdots r_pc_1\cdots c_q)} \\
		=& {\rm Tr}_{r_{p+1}\cdots r_k c_{q+1}\cdots c_{N-k}} \left(\rho\right) \\
		=& {\rm Tr}_{r_{p+1}\cdots r_k c_{q+1}\cdots c_{N-k}} \left(\sum_ip_i \rho^{(l_1\cdots l_k)}_{i} \otimes \rho^{(l_{k+1}\cdots l_N)}_{i}\right) \\
		=& \sum_ip_i {\rm Tr}_{r_{p+1}\cdots r_k}\left(\rho^{(l_1\cdots l_k)}_{i}\right) \otimes {\rm Tr}_{c_{q+1}\cdots c_{N-k}}\left(\rho^{(l_{k+1}\cdots l_N)}_{i}\right) \\
		=& \sum_ip_i \rho^{(r_1\cdots r_p)}_{i} \otimes \rho^{(c_1\cdots c_q)}_{i}.
	\end{aligned}
\end{equation}

By the generalized Bloch representation in Eq.~\eqref{generalized bloch representation n}, $\rho^{(r_1\cdots r_p)}_{i}$, and $\rho^{(c_1\cdots c_q)}_{i}$ can be expanded as
\begin{equation}\label{equation: bloch representation rho^l_r}
	\begin{aligned}
		& \rho^{(r_1\cdots r_p)}_{i} \\
		=& \frac{1}{d_{r_1}\cdots d_{r_p}}\left(\mathbf{I}_{d_{r_1}}^{(r_1)}\otimes\cdots\otimes \mathbf{I}_{d_{r_p}}^{(r_p)} + \cdots\right. \\
		& \left.+ \sum_{i_{r_1},\cdots,i_{r_p}=1}^{d_{r_1}^2-1,\cdots,d_{r_p}^2-1} t_{i_{r_1}\cdots i_{r_p}}^{(i,r_1\cdots r_p)} \mathbf{G}_{i_{r_1}}^{(r_1)}\otimes \cdots \otimes \mathbf{G}_{i_{r_p}}^{(r_p)} \right),
	\end{aligned}
\end{equation}
\begin{equation}\label{equation: bloch representation rho^l_s}
	\begin{aligned}
		& \rho^{(c_1\cdots c_q)}_{i} \\
		=& \frac{1}{d_{c_1}\cdots d_{c_q}}\left(\mathbf{I}_{d_{c_1}}^{(c_1)}\otimes\cdots\otimes \mathbf{I}_{d_{c_q}}^{(c_q)} + \cdots \right.\\
		& \left.+ \sum_{i_{c_1},\cdots,i_{c_q}=1}^{d_{c_1}^2-1,\cdots,d_{c_q}^2-1} t_{i_{c_1}\cdots i_{c_q}}^{(i,c_1\cdots c_q)} \mathbf{G}_{i_{c_1}}^{(c_1)}\otimes \cdots \otimes \mathbf{G}_{i_{c_q}}^{(c_q)} \right),
	\end{aligned}
\end{equation}
where $\mathcal{T}^{(r_1\cdots r_p)}_i=\left(t_{i_{r_1}\cdots i_{r_p}}^{(i,r_1\cdots r_p)}\right)$ and $\mathcal{T}^{(c_1\cdots c_q)}_i=\left(t_{i_{c_1}\cdots i_{c_q}}^{(i,c_1\cdots c_q)}\right)$ are the correlation tensors of $\rho^{(r_1\cdots r_p)}_{i}$ and $\rho^{(c_1\cdots c_q)}_{i}$, respectively.
Substituting Eq.~\eqref{equation: bloch representation rho^l_r} and~\eqref{equation: bloch representation rho^l_s} into Eq.~\eqref{equation: rho_rs}, we have
\begin{equation}\label{equation: rho_rs = sum rho_r rho_s}
	\begin{aligned}
		& \rho^{(r_1\cdots r_pc_1\cdots c_q)} \\
		=& \frac{1}{d_{r_1}\cdots d_{r_p}d_{c_1}\cdots d_{c_q}} \sum_ip_i\\
		& \cdot\left(\mathbf{I}_{d_{r_1}}^{(r_1)}\otimes\cdots\otimes \mathbf{I}_{d_{r_p}}^{(r_p)}\otimes \mathbf{I}_{d_{c_1}}^{(c_1)}\otimes\cdots\otimes \mathbf{I}_{d_{c_q}}^{(c_q)} + \cdots \right.\\
		& + \sum_{i_{r_1},\cdots,i_{r_p}=1}^{d_{r_1}^2-1,\cdots,d_{r_p}^2-1} \sum_{i_{c_1},\cdots,i_{c_q}=1}^{d_{c_1}^2-1,\cdots,d_{c_q}^2-1} t_{i_{r_1}\cdots i_{r_p}}^{(i,r_1\cdots r_p)}t_{i_{c_1}\cdots i_{c_q}}^{(i,c_1\cdots c_q)} \\
		& \left.\cdot\mathbf{G}_{i_{r_1}}^{(r_1)}\otimes \cdots \otimes \mathbf{G}_{i_{r_p}}^{(r_p)} \otimes \mathbf{G}_{i_{c_1}}^{(c_1)}\otimes \cdots \otimes \mathbf{G}_{i_{c_q}}^{(c_q)}\right).
	\end{aligned}
\end{equation}

By the generalized Bloch representation of $\rho$ in Eq.~\eqref{generalized bloch representation n}, $\rho_{r_1\cdots r_pc_1\cdots c_q}$ can be computed as follows,
\begin{equation}\label{equation: bloch representation rho_rs}
	\begin{aligned}
		& \rho^{(r_1\cdots r_pc_1\cdots c_q)} \\
		=& {\rm Tr}_{r_{p+1}\cdots r_kc_{q+1}\cdots c_{N-k}}\left(\rho\right) \\
		=& \frac{1}{d_{r_1}\cdots d_{r_p}d_{c_1}\cdots d_{c_q}} \\
		& \cdot \left(\mathbf{I}_{d_{r_1}}^{(r_1)}\otimes\cdots\otimes \mathbf{I}_{d_{r_p}}^{(r_p)}\otimes \mathbf{I}_{d_{c_1}}^{(c_1)}\otimes\cdots\otimes \mathbf{I}_{d_{c_q}}^{(c_q)} + \cdots \right.\\
		& + \sum_{i_{r_1},\cdots,i_{r_p}=1}^{d_{r_1}^2-1,\cdots,d_{r_p}^2-1} \sum_{i_{c_1},\cdots,i_{c_q}=1}^{d_{c_1}^2-1,\cdots,d_{c_q}^2-1} t_{i_{r_1}\cdots i_{r_p}i_{c_1}\cdots i_{c_q}}^{(r_1\cdots r_pc_1\cdots c_q)} \\
		& \left.\cdot \mathbf{G}_{i_{r_1}}^{(r_1)}\otimes \cdots \otimes \mathbf{G}_{i_{r_p}}^{(r_p)} \otimes \mathbf{G}_{i_{c_1}}^{(c_1)}\otimes \cdots \otimes \mathbf{G}_{i_{c_q}}^{(c_q)}\right),
	\end{aligned}
\end{equation}
where $\mathcal{T}^{(r_1\cdots r_pc_1\cdots c_q)}=\left(t_{i_{r_1}\cdots i_{r_p}i_{c_1}\cdots i_{c_q}}^{(r_1\cdots r_pc_1\cdots c_q)}\right)$ is the correlation tensor of $\rho^{(r_1\cdots r_pc_1\cdots c_q)}$.

Comparing Eq.~\eqref{equation: rho_rs = sum rho_r rho_s} with Eq.~\eqref{equation: bloch representation rho_rs}, we see that
\begin{equation*}
	\begin{gathered}
		t_{i_{r_1}\cdots i_{r_p}i_{c_1}\cdots i_{c_q}}^{(r_1\cdots r_pc_1\cdots c_q)} = \sum_ip_i t_{i_{r_1}\cdots i_{r_p}}^{(i,r_1\cdots r_p)}t_{i_{c_1}\cdots i_{c_q}}^{(i,c_1\cdots c_q)},
	\end{gathered}
\end{equation*}
which implies that $\mathcal{T}^{(r_1\cdots r_pc_1\cdots c_q)} = \sum_i p_i \mathcal{T}^{(r_1\cdots r_p)}\circ\mathcal{T}^{(c_1\cdots c_q)}$.
By Theorem~\ref{theorem: relation between matrix unfolding and outer product}, we have
\begin{equation*}
	\mathbf{T}^{(r_1\cdots r_pc_1\cdots c_q)}_{\left(R,n;C,m\right)} = \sum_ip_i {\rm vec}_{n}\left(\mathcal{T}^{(r_1\cdots r_p)}_{i}\right) {\rm vec}_{m}\left(\mathcal{T}^{(c_1\cdots c_q)}_i\right)^{\rm T}.
\end{equation*}
\end{proof}

\subsection{Proof of Theorem~\ref{lemma: multipartite separability criterion}}\label{section: proof of theorem 4}
\begin{proof}
	Since the state $\rho$ is biseparable under the bipartition $l_1\cdots l_k|l_{k+1}\cdots l_N$, it follows that $\rho$ can be expressed as a convex combination of product sates,
	\begin{equation}\label{equation: convex combination multipartite}
		\rho = \sum_i p_i \rho^{(l_1\cdots l_k)}_{i} \otimes \rho^{(l_{k+1}\cdots l_{N})}_{i},
	\end{equation}
	where $0\leq p_i \leq 1$, $\sum_i p_i =1$, $\rho^{(l_1\cdots l_k)}_{i}$ and $\rho^{(l_{k+1}\cdots l_{N})}_{i}$ are pure states in $\mathcal{H}_{d_{l_1}} \otimes \cdots\otimes\mathcal{H}_{d_{l_k}}$ and $\mathcal{H}_{d_{l_{k+1}}}\otimes \cdots \otimes \mathcal{H}_{d_{l_{N}}}$, respectively. Then, we have\\
	\resizebox{0.5\textwidth}{!}{$
		\begin{aligned}
			& \left\|\mathcal{M}^{(l_1\cdots l_k|l_{k+1}\cdots l_{N})}_{\bm{u},\bm{v},\bm{\alpha},\bm{\beta}}\left(\rho\right) \right\|_{\rm tr} \\
			=& \left\| \begin{pmatrix}
				\bm{u}\bm{v}^{\rm T} & \bm{u}\left(\bm{\beta}\otimes{\rm vec}_{m}\left(\mathcal{T}^{(l_{k+1}\cdots l_{N})}\right)\right)^{\rm T} \\
				\left(\bm{\alpha}\otimes{\rm vec}_{n}\left(\mathcal{T}^{(l_{1}\cdots l_{k})}\right)\right)\bm{v}^{\rm T} & \bm{\alpha}\bm{\beta}^{\rm T}\otimes \mathbf{T}^{(l_{1}\cdots l_{k}  l_{k+1}\cdots l_{N})}_{\left(R,n;C,m\right)}
			\end{pmatrix}\right\|_{\rm tr}\\
			\leq& \sum_ip_i\left\| \begin{pmatrix}
				\bm{u}\bm{v}^{\rm T} &  \bm{u}\left(\bm{\beta}\otimes{\rm vec}_{m}\left(\mathcal{T}_i^{(l_{i_1}\cdots l_{i_j})}\right)\right)^{\rm T} \\
				\left(\bm{\alpha}\otimes{\rm vec}_{n}\left(\mathcal{T}_i^{(l_1\cdots l_k)}\right)\right)\bm{v}^{\rm T} & \bm{\alpha}\bm{\beta}^{\rm T}\otimes \left({\rm vec}_{n}\left(\mathcal{T}^{(l_1\cdots l_k)}_i\right) {\rm vec}_{m}^{\rm T}\left(\mathcal{T}^{(l_{k+1}\cdots l_{N})}_i\right)\right)
			\end{pmatrix}\right\|_{\rm tr}\\
			\leq& \sum_ip_i\left\| \begin{pmatrix}
				\bm{u} \\ \left(\bm{\alpha}\otimes{\rm vec}_{n}\left(\mathcal{T}_i^{(l_1\cdots l_k)}\right)\right)
			\end{pmatrix}\right\|_{\rm tr}
			\left\|\begin{pmatrix}
				\bm{v}^{\rm T} & \left(\bm{\beta}\otimes{\rm vec}_{m}\left(\mathcal{T}_i^{(l_{k+1}\cdots l_{N})}\right)\right)^{\rm T}
			\end{pmatrix}\right\|_{\rm tr}\\
			\leq& \sqrt{ \left\|\bm{u}\right\|_F^2 + \left\|\bm{\alpha}\right\|_F^2 \mathfrak{m}^{(l_1\cdots l_k)}} \sqrt{\left\|\bm{v}\right\|_F^2 +\left\|\bm{\beta}\right\|_F^2 \mathfrak{m}^{(l_{k+1}\cdots l_{N})}},
		\end{aligned}
		$}\\
	where the second equality holds due to Lemma~\ref{lemma: property of correlation tensor n} and the last inequality holds due to Lemma~\ref{lemma: upper bound of correlation tensor n-paritite}. 
\end{proof}

\subsection{Proof of Theorem~\ref{theorem: genuine n-partite entangled}}\label{section: proof of theorem 5}
\begin{proof}
	Suppose that $\rho$ is biseparable. Then it can be expressed as
	\begin{equation*}
		\begin{aligned}
			\rho = \sum_{k=1}^{N-1} \sum_{\substack{\left\{l_1,\cdots,l_N\right\}=[N]\\l_1<\cdots<l_k\\l_{k+1}<\cdots<l_N\\l_1<l_{k+1}}} p_{l_1\cdots l_N} \sum_i p_i^{(l_1\cdots l_k|l_{k+1}\cdots l_N)}& \\
			\rho^{(l_1\cdots l_k)}_{i}\otimes\rho^{(l_{k+1}\cdots l_N)}_{i}&,
		\end{aligned}
	\end{equation*}
	where $0\leq p_i^{(l_1\cdots l_k|l_{k+1}\cdots l_N)}
	\leq1$,
	\begin{equation*}
		\begin{aligned}
			\sum_{k=1}^{N-1}\sum_{\substack{\left\{l_1,\cdots,l_N\right\}=[N]\\l_1<\cdots<l_k\\l_{k+1}<\cdots<l_N\\l_1<l_{k+1}}} p_{l_1\cdots l_N} = 1, \quad \sum_ip_i^{(l_1\cdots l_k|l_{k+1}\cdots l_N)}=1.
		\end{aligned}
	\end{equation*}
	By Theorem~\ref{lemma: multipartite separability criterion}, we obtain that
	\begin{equation*}
		\begin{aligned}
			& M^{(1\cdots N)}(\rho) \\
			=& \frac{1}{2^{N-1}-1}\sum_{k=1}^{N-1}\sum_{\substack{\left\{l_1,\cdots,l_N\right\}=[n]\\l_1<\cdots<l_k\\l_{k+1}<\cdots<l_N\\l_1<l_{k+1}}}\left\|\mathcal{M}_{\bm{u},\bm{v},\bm{\alpha},\bm{\beta}}^{(l_1\cdots l_k|l_{k+1}\cdots l_N)}(\rho)\right\|_{\rm tr}\\
			\leq& \frac{1}{2^{N-1}-1}\sum_{k=1}^{N-1}\sum_{\substack{\left\{l_1,\cdots,l_N\right\}=[n]\\l_1<\cdots<l_k\\l_{k+1}<\cdots<l_N\\l_1<l_{k+1}}} \mathfrak{m}_{\bm{u},\bm{v},\bm{\alpha},\bm{\beta}}^{(l_1\cdots l_k|l_{k+1}\cdots l_N)} \\
			\leq& \frac{1}{2^{N-1}-1} \underbrace{\left(M_0^{(1\cdots N)} +\cdots+ M_0^{(1\cdots N)}\right)}_{2^{N-1}-1}\\
			=& M_0^{(12\cdots N)},
		\end{aligned}
	\end{equation*}
	which is a contradiction. Therefore, $\rho$ is genuine $N$-partite entangled.
\end{proof}

\subsection{Proof of Lemma~\ref{lemma: property of correlation tensor n fully separable}}\label{section: proof of lemma 4}
\begin{proof}
The reduced state $\rho^{(r_1\cdots r_k)}$ can be computed by applying the partial trace to $\rho$, 
\begin{equation}\label{equation: rho r1 rk}
	\begin{aligned}
		\rho^{(r_1\cdots r_k)} =& {\rm Tr}_{[N]\setminus \{r_1,\cdots,r_k\}} \left(\rho\right) \\
		=& {\rm Tr}_{[N]\setminus \{r_1,\cdots,r_k\}} \left(\sum_ip_i\rho^{(1)}_i \otimes \cdots\otimes\rho^{(N)}_i\right) \\
		=& \sum_ip_i\rho^{(r_1)}_i \otimes \cdots\otimes\rho^{(r_k)}_i.
	\end{aligned}
\end{equation}

By the generalized Bloch representation in Eq.~\eqref{generalized bloch representation}, $\rho^{(r_l)}_i$ can be expanded as
\begin{equation}\label{equation: correlation tensor of rho^k_i}
	\rho^{(r_l)}_i = \frac{1}{d_{r_l}} \left(\mathbf{I}_{d_{r_l}}^{(r_l)} + \sum_{i_{r_l}=1}^{d_{r_l}^2-1}t^{(i,r_l)}_{i_{r_l}}\mathbf{G}^{(r_l)}_{i_{r_l}}\right),
\end{equation}
where $\mathcal{T}^{(r_l)}_i = \left(t^{(i,r_l)}_{i_{r_l}}\right) \in\mathbb{C}^{d_{r_l}}$ is the correlation tensor of $\rho^{(r_l)}_i$, $l\in[k]$. Substituting Eq.~\eqref{equation: correlation tensor of rho^k_i} into Eq.~\eqref{equation: rho r1 rk}, we obtain
\begin{equation}\label{equation: expression of rho r1 rk}
	\begin{aligned}
		& \rho^{(r_1\cdots r_k)} \\
		=& \frac{1}{d_{r_1}\cdots d_{r_k}} \sum_ip_i \left(\mathbf{I}_{d_{r_1}}^{(r_1)} + \sum_{i_{r_1}=1}^{d_{r_1}^2-1}t^{(i,r_1)}_{i_{r_1}}\mathbf{G}^{(r_1)}_{i_{r_1}}\right) \otimes\cdots \\
		&\otimes \left(\mathbf{I}_{d_{r_k}}^{(r_k)} + \sum_{i_{r_k}=1}^{d_{r_k}^2-1}t^{(i,r_k)}_{i_{r_k}}\mathbf{G}^{(r_k)}_{i_{r_k}}\right) \\
		=& \frac{1}{d_{r_1}\cdots d_{r_k}} \sum_ip_i \left(\mathbf{I}_{d_{r_1}}^{(r_1)}\otimes\cdots\otimes\mathbf{I}_{d_{r_k}}^{(r_k)}+ \cdots  \right.\\
		&\left.+ \sum_{i_{r_1}=1}^{d_{r_1}^2-1}\cdots\sum_{i_{r_k}=1}^{d_{r_k}^2-1}t^{(i,r_1)}_{i_{r_1}}\cdots t^{(i,r_k)}_{i_{r_k}}\mathbf{G}^{(r_1)}_{i_{r_1}}\otimes\cdots\otimes\mathbf{G}^{(r_k)}_{i_{r_k}} \right).
	\end{aligned}
\end{equation}

By the generalized Bloch representation in Eq.~\eqref{generalized bloch representation n}, $\rho^{(r_1\cdots r_k)}$ can be computed as follows,
\begin{equation}\label{equation: generalized block representation rho r1 rk}
	\begin{aligned}
		\rho^{(r_1\cdots r_k)} =& {\rm Tr}_{[N]\setminus \{r_1,\cdots,r_k\}} \left(\rho\right) \\
		=& \frac{1}{d_{r_1}\cdots d_{r_k}} \left(\mathbf{I}_{d_{r_1}}^{(r_1)}\otimes\cdots\otimes\mathbf{I}_{d_{r_k}}^{(r_k)} + \cdots \right.\\
		& \left. + \sum_{i_{r_1}=1}^{d_{r_1}^2-1}\cdots\sum_{i_{r_k}=1}^{d_{r_k}^2-1}t^{(r_1\cdots r_k)}_{i_{r_1}\cdots i_{r_k}} \mathbf{G}^{(r_1)}_{i_{r_1}}\otimes\cdots\otimes\mathbf{G}^{(r_k)}_{i_{r_k}} \right).
	\end{aligned}
\end{equation}

Comparing Eq.~\eqref{equation: expression of rho r1 rk} and Eq.~\eqref{equation: generalized block representation rho r1 rk}, we have $t^{(r_1\cdots r_k)}_{i_{r_1}\cdots i_{r_k}} = \sum_ip_i t^{(i,r_1)}_{i_{r_1}}\cdots t^{(i,r_k)}_{i_{r_k}}$,
which implies that 
\begin{equation*}
	\mathcal{T}^{(r_1\cdots r_k)} = \sum_ip_i \mathcal{T}^{(r_1)}_i\circ\cdots\circ\mathcal{T}^{(r_k)}_i.
\end{equation*}
\end{proof}

\subsection{Proof of Corollary~\ref{theorem: fully separability}}\label{section: proof of theorem 6}
\begin{proof}
	Since $\rho$ is fully separable, $\rho$ is biseparable under any bipartition $r_1\cdots r_k|c_1\cdots c_{N-k}$, that is, 
	\begin{equation*}
		\rho = \sum_ip_i \rho^{(1)}_i \otimes\cdots\otimes \rho^{(N)}_i = \sum_ip_i \rho^{(r_1\cdots r_k)}_i \otimes \rho^{(c_1\cdots c_{N-k})}_i,
	\end{equation*}
	where $0\leq p_i\leq 1$, $\sum_ip_i=1$, $\rho^{(l)}_i$ is the pure state for the subsystem $\mathcal{H}_{d_l}$, $l\in[N]$, $\rho^{(r_1\cdots r_k)}_i$ and $\rho^{(c_1\cdots c_{N-k})}_i$ are pure states for the subsystems $\mathcal{H}_{d_{r_1}}\otimes\cdots\otimes\mathcal{H}_{r_k}$ and $\mathcal{H}_{d_{c_1}}\otimes\cdots\otimes\mathcal{H}_{c_{N-k}}$, respectively. By Lemma~\ref{lemma: property of correlation tensor n fully separable}, it holds that
	\begin{equation*}
		\begin{gathered}
			\mathcal{T}^{(r_1\cdots r_k)} = \sum_ip_i \mathcal{T}^{(r_1)}_i\circ\cdots\circ\mathcal{T}^{(r_k)}_i,\\
			\mathcal{T}^{(c_1\cdots c_{N-k})} = \sum_ip_i \mathcal{T}^{(c_1)}_i\circ\cdots\circ\mathcal{T}^{(c_{N-k})}_i,
		\end{gathered}
	\end{equation*}
	where $\mathcal{T}^{(r_1\cdots r_k)}$, $\mathcal{T}^{(c_1\cdots c_{N-k})}$, and $\mathcal{T}^{(k)}_i$ are the correlation tensors of $\rho^{(r_1\cdots r_k)}$, $\rho^{(c_1\cdots c_{N-k})}$, and $\rho^{(l)}_i$, respectively, $l\in[N]$. By Lemma~\ref{lemma: bloch vector upper bound}, we have $\left\|\mathcal{T}^{(l)}_i\right\|_F^2\leq\frac{d_l^2-d_l}{\kappa_l}$, $l\in[N]$. 
	
	From the definition of the trace norm of tensors, we have
	\begin{equation*}
		\begin{aligned}
			&\left\|\mathcal{T}^{(1\cdots N)}\right\|_{\rm tr} \\
			=& \max_{\substack{R\cup C=[N], R\cap C=\emptyset\\ \left|R\right|=k>0,\left|C\right|=N-k>0\\ n\in[k],m\in[N-k]}}\left\{\left\|\mathbf{T}^{(r_1\cdots r_k c_1\cdots c_{N-k})}_{(R,n;C,m)}\right\|_{\rm tr}\right\} \\
			=& \max_{\substack{R\cup C=[N], R\cap C=\emptyset\\ \left|R\right|=k>0,\left|C\right|=N-k>0\\ n\in[k],m\in[N-k]}}\left\{ \right.\\
			&\left.\left\| \sum_ip_i {\rm vec}_n\left(\mathcal{T}^{(r_1\cdots r_k)}_i\right) {\rm vec}_m\left(\mathcal{T}^{(c_1\cdots c_{N-k})}_i\right)^{\rm T} \right\|_{\rm tr}\right\} \\
			=& \max_{\substack{R\cup C=[N], R\cap C=\emptyset\\ \left|R\right|=k>0,\left|C\right|=N-k>0\\ n\in[k],m\in[N-k]}}\left\{ \right.\\
			&\left\| \sum_ip_i \left(\mathcal{T}^{(r_{n+1})}_i\otimes\cdots\otimes\mathcal{T}^{(r_{k})}_i\otimes\mathcal{T}^{(r_{1})}_i\otimes\cdots\otimes\mathcal{T}^{(r_n)}_i\right)\right. \\
			& \left.\left. \left(\mathcal{T}^{(c_{m+1})}_i\otimes\cdots\otimes\mathcal{T}^{(c_{N-k})}_i\otimes\mathcal{T}^{(c_{1})}_i\otimes\cdots\otimes\mathcal{T}^{(c_m)}_i\right)^{\rm T} \right\|_{\rm tr}\right\} \\
			\leq& \max_{\substack{R\cup C=[N], R\cap C=\emptyset\\ \left|R\right|=k>0,\left|C\right|=N-k>0\\ n\in[k],m\in[N-k]}}\left\{ \sum_ip_i \prod_{l=1}^N\left\|\mathcal{T}^{(l)}_i\right\|_{\rm tr}\right\} \\
			\leq& \prod_{l=1}^N \sqrt{\frac{d_l^2-d_l}{\kappa_l}}.
		\end{aligned}
	\end{equation*}
\end{proof}

\section{Upper bound of the Frobenius norm of correlation tensors}\label{section: upper bound of frobenius norm of correlation tensor}
In Ref.~\cite{zhao2022detection,zhang2023improved}, two upper bounds on the Frobenius norm of the correlation tensor were established, as detailed below.
\begin{lemma}[Lemma 1 and 3 in Ref.~\cite{zhao2022detection}]\label{lemma: upper bound of correlation tensor zhao}
	Let $\rho\in\mathcal{H}_{d_1}\otimes\mathcal{H}_{d_2}\otimes\cdots\otimes\mathcal{H}_{d_N}$ be an $N$-partite pure quantum state. Then its correlation tensor of the generalized Block representation under the generalized Pauli operator satisfies $ \left\|\mathcal{T}^{(1\cdots N)}\right\|_F^2 \leq \mathfrak{m}^{(1\cdots N)}_{GPauli}$, where
	\begin{equation*}
		\mathfrak{m}^{(1\cdots N)}_{GPauli} = \begin{cases}
			d_1-1, & \mbox{if } N=1, \\
			\frac{d_1\cdots d_N\left(N-1-\sum_{i=1}^N\frac{1}{d_i^2}\right)+1}{N-1}, & \mbox{if } N\geq2.
		\end{cases}
	\end{equation*}
\end{lemma}
\begin{lemma}[Lemma 1, 2, and 3 in Ref.~\cite{zhang2023improved}]\label{lemma: upper bound of correlation tensor zhang}
	Let $\rho\in\mathcal{H}_{d_1}\otimes\mathcal{H}_{d_2}\otimes\cdots\otimes\mathcal{H}_{d_N}$ $(N\geq3)$ be an $n$-partite pure quantum state, and $d=\max_{i\in[N]}\left\{d_i\right\}$. If $\frac{d_1\cdots d_N}{d^2}\geq 1$, then its correlation tensor of the generalized Block representation under the Weyl operators satisfies $\left\|\mathcal{T}^{(1\cdots N)}\right\|_F ^2 \leq \mathfrak{m}^{(1\cdots N)}_{Weyl}$, where
	\begin{equation*}
		\begin{aligned}
			& \mathfrak{m}^{(1\cdots N)}_{Weyl} \\
			=& \begin{cases}
				d_1-1, & \mbox{if } N=1,\\
				d_1d_2\left(1 - \frac{1}{d^2}\right), & \mbox{if } N=2, \\
				\makecell{\scriptsize\begin{aligned}
					&d_1\cdots d_N - \frac{d_1\cdots d_N}{N-1}\sum_{i=1}^{N}\frac{1}{d_i^2} + \frac{1}{N-1} \\ 
					&\quad+ \frac{1}{N-2}\left(\frac{N}{N-1} - \frac{d_1\cdots d_N}{N-1}\sum_{i=1}^{N}\frac{1}{d_i^2}\right),
				\end{aligned}} & \mbox{if } N\geq3, \frac{d_1\cdots d_N}{d^2}\geq1.
			\end{cases}
		\end{aligned}
	\end{equation*}
\end{lemma}

Now, we discuss the relation between our upper bound in Lemma~\ref{lemma: upper bound of correlation tensor n-paritite} and those in Lemma~\ref{lemma: upper bound of correlation tensor zhao} and~\ref{lemma: upper bound of correlation tensor zhang}. If the traceless orthogonal basis in this paper is chosen as the generalized Pauli operator or the Weyl operator, then we have $\kappa_i=d_i$, $i\in[N]$. In this case, the upper bound $\mathfrak{m}^{(1\cdots N)}$ in Lemma~\ref{lemma: upper bound of correlation tensor n-paritite} reduces to 
\begin{equation*}
	\begin{aligned}
		& \bar{\mathfrak{m}}^{(1\cdots N)} \\
		=& \begin{cases}
			d_1 - 1, & \mbox{if } N=1, \\
			\makecell{\scriptsize\begin{aligned}
				&d_1\cdots d_N + \frac{1}{N-1} \\
				&\quad- \frac{d_1\cdots d_N}{N-1} \sum_{i=1}^N \frac{1}{d_i^2},
			\end{aligned}} & \mbox{if } N=2 \mbox{ or } N\geq3 , \frac{d_1\cdots d_N}{d^2} < 1, \\
			\makecell{\scriptsize\begin{aligned}
				&d_1\cdots d_N + \frac{2}{N-2} \\
				&\quad- \frac{d_1\cdots d_N}{N-2} \sum_{i=1}^N \frac{1}{d_i^2},
			\end{aligned}} & \mbox{if } N\geq3, \frac{d_1\cdots d_N}{d^2} \geq 1.
		\end{cases}
	\end{aligned}
\end{equation*}

When $N=1$. The three upper bounds are the same.

When $N=2$. The upper bound of Lemma~\ref{lemma: upper bound of correlation tensor zhao} in Ref.~\cite{zhao2022detection} is the same as ours. Suppose that $d_1\geq d_2$. Then we have $\bar{\mathfrak{m}}^{(1\cdots N)} - \mathfrak{m}^{(1\cdots N)}_{Weyl} \leq 0$, which implies that our upper bound is sharper than that of Lemma~\ref{lemma: upper bound of correlation tensor zhang} in Ref.~\cite{zhang2023improved}.

When $N\geq3$, $\frac{d_1\cdots d_N}{d^2}<1$. The upper bound of Lemma~\ref{lemma: upper bound of correlation tensor zhao} in Ref.~\cite{zhao2022detection} is the same as ours. 

When $N\geq3$, $\frac{d_1\cdots d_N}{d^2}\geq1$. We have $\bar{\mathfrak{m}}^{(1\cdots N)} - \mathfrak{m}^{(1\cdots N)}_{GPauli} \leq 0$ and $\bar{\mathfrak{m}}^{(1\cdots N)} - \mathfrak{m}^{(1\cdots N)}_{Weyl} = 0$,
which implies that our upper bound is sharper than that of Lemma~\ref{lemma: upper bound of correlation tensor zhao} in Ref.~\cite{zhao2022detection}, and the upper bound of Lemma~\ref{lemma: upper bound of correlation tensor zhang} in Ref.~\cite{zhang2023improved} is the same as ours.

\bibliographystyle{unsrtnat}
\bibliography{references}

\begin{thebibliography}{48}
\providecommand{\natexlab}[1]{#1}
\providecommand{\url}[1]{\texttt{#1}}
\expandafter\ifx\csname urlstyle\endcsname\relax
  \providecommand{\doi}[1]{doi: #1}\else
  \providecommand{\doi}{doi: \begingroup \urlstyle{rm}\Url}\fi

\bibitem[Nielsen and Chuang(2010)]{nielsen2010quantum}
M.~A. Nielsen and I.~L. Chuang.
\newblock \emph{Quantum Computation and Quantum Information}.
\newblock Cambridge University, 2010.

\bibitem[Horodecki et~al.(2009{\natexlab{a}})Horodecki, Horodecki, Horodecki,
  and Horodecki]{horodecki2009quantum}
R.~Horodecki, P.~Horodecki, M.~Horodecki, and K.~Horodecki.
\newblock Quantum entanglement.
\newblock \emph{Reviews of Modern Physics}, 81:\penalty0 865,
  2009{\natexlab{a}}.

\bibitem[Huo et~al.(2026)Huo, Wang, Qu, Xing, Zhan, Huang, Gao, Xiao, and
  Xue]{huo2026sequential}
B.~Huo, K.~Wang, D.~Qu, J.~Xing, X.~Zhan, X.~Huang, H.~Gao, L.~Xiao, and
  P.~Xue.
\newblock Sequential sharing of quantum nonlocality via projective
  measurements.
\newblock \emph{Physical Review Letters}, 136:\penalty0 100201, 2026.

\bibitem[Li et~al.(2025{\natexlab{a}})Li, Xing, Qu, Gao, Xiao, Liu, Xiao, and
  Xue]{li2025temporal}
Y.~Li, J.~Xing, D.~Qu, H.~Gao, L.~Xiao, J.~Liu, Y.~Xiao, and P.~Xue.
\newblock Temporal asymmetry in entanglement distillation.
\newblock \emph{Physical Review Letters}, 135:\penalty0 170801,
  2025{\natexlab{a}}.

\bibitem[DiVincenzo(1995)]{divincenzo1995quantum}
D.P. DiVincenzo.
\newblock Quantum computation.
\newblock \emph{Science}, 1995.

\bibitem[Bennett et~al.(1993)Bennett, Brassard, Crépeau, Jozsa, Peres, and
  Wootters]{bennett1993teleporting}
C.H. Bennett, G.~Brassard, C.~Crépeau, R.~Jozsa, A.~Peres, and W.K. Wootters.
\newblock Teleporting an unknown quantum state via dual classical and
  einstein-podolsky-rosen channels.
\newblock \emph{Physical Review Letters}, 70, 1993.

\bibitem[Xu et~al.(2020)Xu, Ma, Zhang, Lo, and Pan]{xu2020secure}
F.H. Xu, X.F. Ma, Q.~Zhang, H.K. Lo, and J.W. Pan.
\newblock Secure quantum key distribution with realistic devices.
\newblock \emph{Reviews of Modern Physics}, 92:\penalty0 025002, 2020.

\bibitem[Xing et~al.(2024)Xing, Li, Qu, Xiao, Fan, Ma, Xue, Bharti, Koh, and
  Xiao]{xing2024teleportation}
J.~Xing, Y.~Li, D.~Qu, L.~Xiao, Z.~Fan, H.~Ma, P.~Xue, K.~Bharti, Dax~E. Koh,
  and Y.~Xiao.
\newblock Teleportation with embezzling catalysts.
\newblock \emph{Communications Physics}, 7:\penalty0 357, 2024.

\bibitem[Li et~al.(2025{\natexlab{b}})Li, Xing, Qu, Xiao, Fan, Zheng, Ma, Xue,
  Bharti, Koh, and Xiao]{li2025communication}
Y.~Li, J.~Xing, D.~Qu, L.~Xiao, Z.~Fan, Z.J. Zheng, H.~Ma, P.~Xue, K.~Bharti,
  Dax~E. Koh, and Y.~Xiao.
\newblock Communication with quantum catalysts.
\newblock \emph{Quantum Science and Technology}, 10\penalty0 (3):\penalty0
  035044, 2025{\natexlab{b}}.

\bibitem[Ekert(1991)]{ekert1991quantum}
A.K. Ekert.
\newblock Quantum cryptography based on bell’s theorem.
\newblock \emph{Physical Review Letters}, 67:\penalty0 661, 1991.

\bibitem[Lloyd(1996)]{lioyd1996universal}
S.~Lloyd.
\newblock Universal quantum simulators.
\newblock \emph{Science}, 273:\penalty0 1073, 1996.

\bibitem[Peres(1996)]{peres1996separability}
A.~Peres.
\newblock Separability criterion for density matrices.
\newblock \emph{Physical Review Letters}, 77:\penalty0 1413, 1996.

\bibitem[Horodecki et~al.(1996)Horodecki, Horodecki, and
  Horodecki]{horodecki1996separability}
M.~Horodecki, P.~Horodecki, and R.~Horodecki.
\newblock Separability of mixed states: necessary and sufficient conditions.
\newblock \emph{Physical Letter A}, 223, 1996.

\bibitem[Li et~al.(2024)Li, Yao, Fei, Fan, and Ma]{li2024quantum}
J.~Li, H.~Yao, S.M. Fei, Z.~Fan, and H.~Ma.
\newblock Quantum entanglement estimation via symmetric-measurement-based
  positive maps.
\newblock \emph{Physical Review A}, 109:\penalty0 052426, 2024.

\bibitem[Rudolph(2003)]{rudolph2003some}
O.~Rudolph.
\newblock Some properties of the computable cross-norm criterion for
  separability.
\newblock \emph{Physical Review A}, 67:\penalty0 032312, 2003.

\bibitem[Rudolph(2005)]{rudolph2005further}
O.~Rudolph.
\newblock Further results on the cross norm criterion for separability.
\newblock \emph{Quantum Information Processing}, 4, 2005.

\bibitem[Chen and Wu(2002)]{chen2002matrix}
K.~Chen and L.A. Wu.
\newblock A matrix realignment method for recognizing entanglement.
\newblock \emph{Quantum Information and Computation}, 3, 2002.

\bibitem[Sun et~al.(2025)Sun, Yao, Fei, and Fan]{sun2025separability}
J.X. Sun, H.M. Yao, S.M. Fei, and Z.B. Fan.
\newblock Separability and lower bounds of quantum entanglement based on
  realignment.
\newblock \emph{Physical Review A}, 111:\penalty0 012415, 2025.

\bibitem[Gühne et~al.(2007)Gühne, Hyllus, Gittsovich, and
  Eisert]{guhne2007covariance}
O.~Gühne, P.~Hyllus, O.~Gittsovich, and J.~Eisert.
\newblock Covariance matrices and the separability problem.
\newblock \emph{Physical Review Letters}, 99:\penalty0 130504, 2007.

\bibitem[de~Vicente(2007)]{devicente2007separability}
J.I. de~Vicente.
\newblock Separability criteria based on the bloch representation of density
  matrices.
\newblock \emph{Quantum Information and Computation}, 7, 2007.

\bibitem[Bu et~al.(2015)Bu, Wang, Sun, and Zhou]{bu2015minimum}
C.~Bu, W.~Wang, L.~Sun, and J.~Zhou.
\newblock Minimum (maximum) rank of sign pattern tensors and sign nonsingular
  tensors.
\newblock \emph{Linear Algebra and its Applications}, 483:\penalty0 101--114,
  2015.

\bibitem[Yao et~al.(2015)Yao, Liu, and Bu]{yao2015tensors}
H.~Yao, L.~Liu, and C.~Bu.
\newblock Tensors product and hyperdeterminant of boundary formats product.
\newblock \emph{Linear Algebra and its Applications}, 483:\penalty0 1--20,
  2015.

\bibitem[Yao et~al.(2016)Yao, Long, Bu, and Zhou]{yao2016singular}
H.~Yao, B.~Long, C.~Bu, and J.~Zhou.
\newblock lk,s-singular values and spectral radius of partially symmetric
  rectangular tensors.
\newblock \emph{Frontiers of Mathematics in China}, 11:\penalty0 605--622,
  2016.

\bibitem[Sun et~al.(2016)Sun, Zheng, Bu, and Wei]{sun2016moore}
L.~Sun, B.~Zheng, C.~Bu, and Y.~Wei.
\newblock Moore–penrose inverse of tensors via einstein product.
\newblock \emph{Linear and Multilinear Algebra}, 64\penalty0 (4):\penalty0
  686--698, 2016.

\bibitem[Sun et~al.(2018)Sun, Zheng, Wei, and Bu]{sun2018generalized}
L.~Sun, B.~Zheng, Y.~Wei, and C.~Bu.
\newblock Generalized inverses of tensors via a general product of tensors.
\newblock \emph{Frontiers of Mathematics in China}, 13:\penalty0 893--911,
  2018.

\bibitem[Shen et~al.(2016)Shen, Yu, Li, and Fei]{shen2016improved}
S.Q. Shen, J.~Yu, M.~Li, and S.M. Fei.
\newblock Improved separability criteria based on bloch representation of
  density matrices.
\newblock \emph{Scientific Reports}, 6:\penalty0 28850, 2016.

\bibitem[Chang et~al.(2018)Chang, Cui, Zhang, and Fei]{chang2018separability}
J.M. Chang, M.Y. Cui, T.G. Zhang, and S.M. Fei.
\newblock Separability criteria based on heisenberg-weyl representation of
  density matrices.
\newblock \emph{Chinese Physics B}, 27:\penalty0 030302, 2018.

\bibitem[Zhao et~al.(2022{\natexlab{a}})Zhao, Yang, Jing, Wang, and
  Fei]{zhao2022detectionofmultipartite}
H.~Zhao, Y.~Yang, N.H. Jing, Z.X. Wang, and S.M. Fei.
\newblock Detection of multipartite entanglement based on heisenberg-weyl
  representation of density matrices.
\newblock \emph{International Journal of Theoretical Physics}, 61\penalty0
  (5):\penalty0 136, 2022{\natexlab{a}}.

\bibitem[Sarbicki et~al.(2020)Sarbicki, Scala, and
  Chruściński]{sarbicki2020family}
G.~Sarbicki, G.~Scala, and D.~Chruściński.
\newblock Family of multipartite separability criteria based on a correlation
  tensor.
\newblock \emph{Physical Review A}, 101:\penalty0 012341, 2020.

\bibitem[Huang et~al.(2022)Huang, Zhang, Zhao, and Jing]{huang2022separability}
X.F. Huang, T.G. Zhang, M.J. Zhao, and N.H. Jing.
\newblock Separability criteria based on the weyl operators.
\newblock \emph{Entropy}, 24:\penalty0 1064, 2022.

\bibitem[Zhu et~al.(2023)Zhu, Wang, Bao, Li, Shen, and Fei]{zhu2023family}
X.N. Zhu, J.~Wang, G.~Bao, M.~Li, S.Q. Shen, and S.M. Fei.
\newblock A family of bipartite separability criteria based on bloch
  representation of density matrices.
\newblock \emph{Quantum Information Processing}, 22:\penalty0 185, 2023.

\bibitem[Huang et~al.(2024)Huang, Zhang, and Jing]{huang2024unifying}
X.F. Huang, T.G. Zhang, and N.H. Jing.
\newblock A unifying separability criterion based on extended correlation
  tensor.
\newblock \emph{Quantum Information Processing}, 23:\penalty0 233, 2024.

\bibitem[Bader and Kolda(2006)]{brett2006algorithm}
Brett~W. Bader and Tamara~G. Kolda.
\newblock Algorithm 862: {MATLAB} tensor classes for fast algorithm
  prototyping.
\newblock \emph{ACM Transactions on Mathematical Software}, 32\penalty0
  (4):\penalty0 635--653, 2006.

\bibitem[Kolda and Bader(2009)]{kolda2009tensor}
Tamara~G. Kolda and Brett~W. Bader.
\newblock Tensor decompositions and applications.
\newblock \emph{SIAM Review}, 51\penalty0 (3):\penalty0 455--500, 2009.

\bibitem[De~Lathauwer et~al.(2000)De~Lathauwer, De~Moor, and
  Vandewalle]{delathauwer2000multilinear}
L.~De~Lathauwer, B.~De~Moor, and J.~Vandewalle.
\newblock A multilinear singular value decomposition.
\newblock \emph{SIAM Journal on Matrix Analysis and Applications}, 21, 2000.

\bibitem[Bennett et~al.(1999)Bennett, DiVincenzo, Mor, Shor, Smolin, and
  Terhal]{bennett1999unextendible}
C.H. Bennett, D.P. DiVincenzo, T.~Mor, P.W. Shor, J.A. Smolin, and B.M. Terhal.
\newblock Unextendible product bases and bound entanglement.
\newblock \emph{Physical Review Letters}, 82:\penalty0 5385, 1999.

\bibitem[Li et~al.(2014)Li, Wang, Fei, and Li-Jost]{li2014quantum}
M.~Li, J.~Wang, S.M. Fei, and X.Q. Li-Jost.
\newblock Quantum separability criteria for arbitrary-dimensional multipartite
  states.
\newblock \emph{Physical Review A}, 89:\penalty0 022325, 2014.

\bibitem[Horodecki(1997)]{horodechi1997separability}
P.~Horodecki.
\newblock Separability criterion and inseparable mixed states with positive
  partial transposition.
\newblock \emph{Physics Letters A}, 232\penalty0 (5):\penalty0 333--339, 1997.

\bibitem[Zhao et~al.(2022{\natexlab{b}})Zhao, Liu, Jing, and
  Wang]{zhao2022detection}
H.~Zhao, Y.Q. Liu, N.H. Jing, and Z.X. Wang.
\newblock Detection of genuine entanglement for multipartite quantum states.
\newblock \emph{Quantum Information Processing}, 21:\penalty0 315,
  2022{\natexlab{b}}.

\bibitem[Zhang et~al.(2023)Zhang, Jing, Zhao, Liu, and Ma]{zhang2023improved}
X.~Zhang, N.H. Jing, H.~Zhao, M.~Liu, and H.T. Ma.
\newblock Improved tests of genuine entanglement for multiqudits.
\newblock \emph{Europhysics Letters}, 143:\penalty0 38002, 2023.

\bibitem[Horodecki et~al.(2009{\natexlab{b}})Horodecki, Horodecki, Horodecki,
  and Horodecki]{horodechi2009quantum}
R.~Horodecki, P.~Horodecki, M.~Horodecki, and K.~Horodecki.
\newblock Quantum entanglement.
\newblock \emph{Reviews of Modern Physics}, 81:\penalty0 865--942,
  2009{\natexlab{b}}.

\bibitem[Hassan and Joag(2008)]{hassan2008separability}
Ali S.~M. Hassan and Pramod~S. Joag.
\newblock Separability criterion for multipartite quantum states based on the
  bloch representation of density matrices.
\newblock \emph{Quantum Information and Computation}, 8\penalty0 (8):\penalty0
  773–790, 2008.

\bibitem[Laskowski et~al.(2011)Laskowski, Markiewicz, Paterek, and
  Żukowski]{laskowki2011correlation}
W.~Laskowski, M.~Markiewicz, T.~Paterek, and M.~Żukowski.
\newblock Correlation-tensor criteria for genuine multiqubit entanglement.
\newblock \emph{Physical Review A}, 84:\penalty0 062305, 2011.

\bibitem[Cornelio and de~Toledo~Piza(2006)]{cornelio2006classification}
Marcio~F. Cornelio and A.~F.~R. de~Toledo~Piza.
\newblock Classification of tripartite entanglement with one qubit.
\newblock \emph{Physical Review A}, 73:\penalty0 032314, 2006.

\bibitem[Gittsovich et~al.(2010)Gittsovich, Hyllus, and
  G\"uhne]{gittsovich2010multipartite}
O.~Gittsovich, P.~Hyllus, and O.~G\"uhne.
\newblock Multiparticle covariance matrices and the impossibility of detecting
  graph-state entanglement with two-particle correlations.
\newblock \emph{Physical Review A}, 82:\penalty0 032306, 2010.

\bibitem[Zhang et~al.(2017)Zhang, Lu, Wang, and Shen]{zhang2017realignment}
Y.H. Zhang, Y.Y. Lu, G.B. Wang, and S.Q. Shen.
\newblock Realignment criteria for recognizing multipartite entanglement of
  quantum states.
\newblock \emph{Quantum Information Processing}, 16\penalty0 (4):\penalty0 106,
  2017.

\bibitem[Shen et~al.(2022)Shen, Chen, Hu, and Li]{shen2022optimization}
S.Q. Shen, L.~Chen, A.W. Hu, and M.~Li.
\newblock Optimization of realignment criteria and its applications for
  multipartite quantum states.
\newblock \emph{Quantum Information Processing}, 21\penalty0 (4):\penalty0 135,
  2022.

\bibitem[Asadian et~al.(2016)Asadian, Erker, and Huber]{asadian2016heisenberg}
A.~Asadian, P.~Erker, and M.~Huber.
\newblock Heisenberg-weyl observables: Bloch vectors in phase space.
\newblock \emph{Physical Review A}, 94:\penalty0 010301, 2016.

\end{thebibliography}

\end{document}